\documentclass[12pt]{iopart}

\usepackage{mathptmx}
\usepackage{amsmath}
\usepackage{etoolbox}

\usepackage{subfigure}
\usepackage[labelformat=parens,labelsep=quad,skip=3pt]{caption}
\usepackage{graphicx}
\begin{document}

\title{Particle creation and annihilation in an exclusion process on networks}

\author{Ankita Gupta \& Arvind Kumar Gupta}

\address{Department of Mathematics, Indian Institute of Technology Ropar, Rupnagar-140001, Punjab, India}
\ead{akgupta@iitrpr.ac.in}
\vspace{10pt}

\begin{abstract}
To mimic the complex transport-like collective phenomena in a man-made or natural system, we study an open network junction model of totally asymmetric simple exclusion process with bulk particle attachment and detachment. The stationary system properties such as particle density, phase transitions and phase diagrams are derived theoretically utilising the mean-field approach. The steady-state phases have been categorized into various sub-classes based upon the phase transitions occurring across the junction. It is found that the number of steady-state phases depends on the number of incoming and outgoing segments at the junction. Further, an increase in the particle non-conserving rates significantly affects the topology of the phase diagram and the number of stationary phases changes in a non-monotonic way. For both the case of equal and unequal incoming and outgoing segments, the critical values of non-conserving rates at which the topology of the phase diagram changes are identified. The theoretical results are validated using extensive Monte Carlo simulations.
\end{abstract}

%
% Uncomment for keywords
%\vspace{2pc}
%\noindent{\it Keywords}: XXXXXX, YYYYYYYY, ZZZZZZZZZ
%
% Uncomment for Submitted to journal title message
%\submitto{\JPA}
%
% Uncomment if a separate title page is required
%\maketitle
% 
% For two-column output uncomment the next line and choose [10pt] rather than [12pt] in the \documentclass declaration
%\ioptwocol
%

\section{Introduction}
Transport phenomena have been observed at almost all levels, ranging from man-made structures such as vehicular motion on road networks \cite{Nagel,Chowdhury} or movement of data packets on the internet to natural structures such as intracellular transport of biological molecular motors along cytoskeletal filaments \cite{Klumpp,Chou}, etc. In the majority of the cases, the stochastic motion of the entities happens on a complex network which may give rise to a traffic jam-like situation. In traffic flow, the appearance of jams can cause pollution and increase the consumption of combustible liquid whereas, in biological transport, the molecular motors clogging leads to certain diseases such as Alzheimer’s disease \cite{goldstein} and some neurodegenerative diseases \cite{hurd1996kinesin}. Generally, in such systems, the entities group together to work as a team, therefore, understanding the macroscopic properties of the stochastic system on a network becomes manageable. One of the distinguishing features of such complex systems is the non-zero continuous current which classifies them into non-equilibrium systems. Therefore, a unified framework can be utilized to explain the critical behaviour of all these processes.

For the comprehensive study of the dynamics on a network, it is essential to understand the stationary properties on an individual lane which have been widely scrutinized using the lattice gas models. In this direction, the Totally asymmetric simple exclusion process (TASEP) has gained a paradigmatic status among the class of driven diffusive lattice models that uncovers the non-trivial facts of stochastic transport systems. It was originally introduced in $1968$, in a more general form, as a theoretical model for analysing the kinetics of biopolymerization \cite{macdonald1969concerning, macdonald1968kinetics}. In a single-lane TASEP model, particles hop stochastically in  a preferred direction subject to excluded volume interactions. The boundary conditions play a crucial role in understanding the dynamics of the system. In contrary to the periodic boundary case, the open boundary conditions display must richer phenomena such as boundary induced phase transitions \cite{Krug,Derrida, Muhuri1}, spontaneous symmetry breaking \cite{schutz2000exactly, Krug, Schutz}, etc. Over the decades, numerous generalizations of TASEP have been developed including particle-particle interactions \cite{Midha, Antal}, multi-lane models \cite{dhiman,Wang,Gupta}, bidirectional transport \cite{Zia, Muhuri, Sharma}, etc. 

The key to rationalizing the stochastic dynamics on a network is to understand the processes at the junction. Junctions can be thought of as locations where traffic changes its route or directions. Several generalisations of the TASEP network with junctions have been employed. For example, the quantitative characterization of  single-lane road which bifurcates into two equivalent branches and subsequently merges again into a single lane has been well examined \cite{brankov2004totally}. This can be seen as a model of two consecutive junctions on a single TASEP segment. In literature, a network of $m$ incoming and $n$ outgoing segments connected via a junction has been labeled as $V(m:n)$. Theoretical investigation of $V(2:1)$ has been explored using mean field approximation and extensive computer simulations \cite{pronina2005theoretical}. Inspired by real phenomenons, the study of junctions was extended for multiple input multiple output systems as well as to multiple junctions \cite{wang2008theoretical, song2009theoretical}. Owing to the extensive body of TASEP, the dynamics of traffic flow on $V(2:2)$ with parallel updating rules has also been studied \cite{ishibashi1996phase, ishibashi2001phase, ishibashi2001phases, jindal2021exclusion}. The behaviour of all possible fourfold junctions ($V(2:2),\ V(1:3),\ V(3:1)$) has been thoroughly investigated with explicit vertex framework \cite{embley2009understanding} and further extended to $V(m:n)$ with interacting particles. Recently, \cite{zhang2019exclusion} study aspects of the temporal evolution in the initial particle density of $V(1:2)$ and $V(2:1)$ junctions. Another variant has shown the multiplex structure of the closed networks affects the global traffic flow in a non-trivial way \cite{shen2020totally,neri2013exclusion}. All these papers have focused on the minimal model of TASEP adopted random updating rules where particles do not interact with the surrounding environment. Such studies are very well suited for understanding stochastic transport on networks such as vehicles on roads or motor proteins on biofilaments, etc. 

Further, various studies have focused on coupling the exclusion processes to a bulk reservoir where attachment/detachment of particles prevails on bulk sites (known as Langmuir kinetics (LK)). The importance of studying TASEP with LK (TASEP-LK) lies in its application to intracellular dynamics, where motor proteins exhibit microscopic reversibility between the cytoplasm and the molecular filaments. This inclusion leads to rich stationary behaviour such as non-constant linear density profiles, localized shocks, and continuous phase transitions \cite{ popkov2003localization,evans2003shock, parmeggiani2004totally, parmeggiani2003phase,Vuijk,Ichiki,Ichiki1}. Closed networks coupled with LK have been studied comprehensively and it is found that the particle non-conserving dynamics affect the steady-state properties of the system significantly \cite{neri2013exclusion,Neri,Neri1}.

As discussed earlier that one-segment open systems display several interesting phenomena as compared to closed counterparts, in this work, we explore a network of open TASEP-LK consisting of $m$ incoming segments connected via a junction to $n$ outgoing segments represented as $V(m:n)$. To explore the overall dynamics of the proposed model, we compute theoretical expressions for particle density, phase diagrams, and phase transitions. Our system dynamics encourage us to answer a few essential queries: (i) Does the number of segments regulates the stationary properties of the system? (ii) How the association-dissociation rates govern the dynamics of the system? The paper is organized as follows. In section \ref{Network Model}, we introduce our model and its governing dynamical rules whereas subsection \ref{one} briefly discusses the results for single-segment TASEP-LK model. In subsection \ref{main}, the framework of mean field  is utilized to successfully capture the steady-state properties of the network for thorough analysis. The phase diagrams are theoretically computed in section \ref{Results Discussions}. Finally, we summarize and conclude in section \ref{conclude}.

\section{\label{Network Model}Network Model}
\begin{figure*}
\centering
\includegraphics{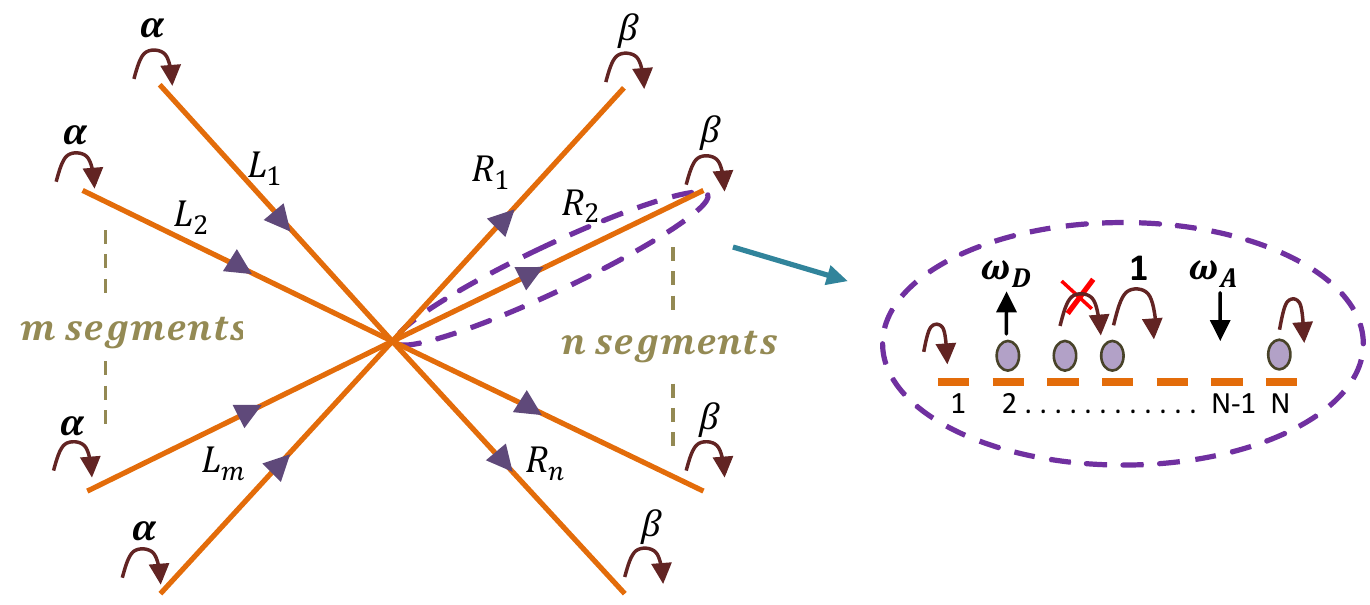}
\caption{Schematic demonstration of the model where $m$ incoming segments and $n$ outgoing segments are connected via a junction. The particle can enter the first vacant site of incoming segments with rate $\alpha$ and exit from the last site of outgoing segments with rate $\beta$. To the right, zoom of a segment is shown where attachment (detachment) of particles can occur in the bulk of each segment with rate $\omega_A\ (\omega_D)$. }
\label{model}
\end{figure*}

To mimic the stochastic transport of particles on a network such as vehicular traffic on a road,  the motion of molecular motors along microtubules, etc, the present work explores the collective dynamics over a complex network which comprises of $m$ independent incoming segments interacting with $n$ outgoing segments at a junction.
We propose a  network $V(m:n)$ of open TASEPs composing of two subsystems: the left $(L)$ subsystem consists of $m$ incoming segments $L_1, L_2,\ldots,L_m$ and the right  $(R)$ subsystem comprises of $n$ outgoing segments $R_1, R_2,\ldots,R_n$ connected via a junction with particle-creation and annihilation (see figure \ref{model}). Each $L_k$ and $R_k$ segment represents a uniform open TASEP consisting of $N$ sites and the complete network can be regarded as a system with $(m+n) N$ sites. The sites $i=1$ and $i=N$ constitute the boundaries whereas $1<i<N$ represents the bulk of an individual segment.  The system is connected to an infinite reservoir of indistinguishable particles which move in a preferred direction (left to right) following a random sequential update rule. Particles are distributed under the hard-core exclusion principle which ensures that not more than one particle can occupy a segment site. It is assumed that neither intra-subsystems nor inter-subsystems particle-particle interactions are permitted directly and the particles of the two subsystems are only allowed to interact at the junction.

\textit{\textbf{Dynamics on an individual segment:}}
 Particles are injected into the system through the first  site $(i=1)$ of each $L_k$ segment with a rate $\alpha$ if empty. A particle from this site can hop with unit rate to the empty neighboring site.
In the bulk, a particle first tries to leave the system with detachment rate $\omega_D$. If it fails, then  it attempts to jumps to the site $i+1$ with unit rate if the target site is empty, obeying the hard-core exclusion principle. Furthermore, if the $i^\text{th}$ site is vacant, then a particle can enter the system with rate $\omega_A$.  A particle finally exits from the last site $(i=N)$ of each $R_k$ segment with rate $\beta$.

\textit{\textbf{Dynamics at  the junction:}}  Particle on the site $i=N$ of any $L_k$ segment, can jump to the site $i=1$ of any $R_k$ segment with equal rate following the hard-core exclusion principle. In case of competition, where more than one particle is available to jump across the junction, then they have equal chances of hopping onto the first site of any of the $R_k$ segments.

It is worthwhile pointing out that, in the absence of LK dynamics, reference \cite{wang2008theoretical} can be considered a specific case of our model with $n=1$ and further for $m=2$ and  $n=1$, the present model reduces to refrence \cite{zhang2019exclusion}. 
\section{\label{theory}Theoretical Description}
In the presented network model, all the individual segments are homogeneous and the system dynamics will entirely be governed by the behaviour of each segment at the steady state. It is worth here to recall the dynamic properties of a single segment homogeneous TASEP with open boundaries coupled with Langmuir kinetics (LK) for random updating rule which has been thoroughly examined in the literature \cite{parmeggiani2003phase,parmeggiani2004totally} and analysed utilizing a very generic approach known as Mean Field Approximation. The mean field approximation assumes that the probability of occupancy of any site is independent of the occupancies of other sites and it also ignores all kinds of interactions in the system.

\subsection{\label{one}
\textit{\textbf{One-segment TASEP with particle-creation and annihilation:}}}

To determine the overall state of the system, the master equation for a one-dimensional segment of $N$ sites has been examined at the steady state \cite{parmeggiani2004totally}. The equation exhibits the temporal evolution of the particle densities on each site of the segment. 
Continuum limit of this system can be obtained by coarse-graining the discrete lattice with lattice constant $\epsilon=1/N$ and rescaling the time as $t'=t/N$. Introduce the rescaled attachment, detachment rates and the binding constant as $\Omega_A=\omega_A N,\ \Omega_D=\omega_D N$ and  $K=\Omega_A/\Omega_D$. The hydrodynamic behaviour is governed by the continuity equation 
\begin{equation}\label{bulklk}
\frac{\partial\rho}{\partial t'} = \partial_x\left(\frac{\epsilon}{2}\partial_x\rho-\rho(1-\rho)\right)+\omega_A(1-\rho)-\omega_D\rho.
\end{equation}
where $x=i/N,\ 0\leq x \leq 1$ denotes the rescaled position variable and $\rho$ gives the average particle density.
We focus on the special case when $\Omega_A=\Omega_D=\Omega$, for which equation (\ref{bulklk})
 at steady state reduces to 
 \begin{equation}
  \frac{\epsilon}{2}\frac{d^2\rho}{dx^2}+(2\rho-1)\frac{d\rho}{dx}+\Omega(1-2\rho)=0
 \end{equation}
 along with the boundary conditions $\rho(0)=\alpha$ and $\rho(1)=1-\beta.$ In the continuum limit $\epsilon \rightarrow 0^+$, it has been predicted \cite{parmeggiani2003phase,parmeggiani2004totally} that the system dynamics is specified by the entrance rate $\alpha$, exit rate $\beta$ and the attachment/detachment rate $\Omega$. The obtained density profiles are piecewise linear and continuously dependent upon $\Omega$. The bulk particle density at the steady state is given by $\Omega x+C$ where the constant $C$ depends on the boundary condition satisfied. Another solution obtained is $\rho_l(x)=1/2$ identical to the Langumir isotherm and also the density of the MC phase of TASEP which remains unaffected by $\Omega$. 
\begin{table}[htb] 
 \caption{Mean-field expressions of the phase boundaries for one-dimensional open TASEP coupled with LK. }
 \begin{center}
 
\begin{tabular}{||l||c||l||}
\hline \hline
Phase & Phase Boundaries & $\rho (x)$\\ \hline \hline
LD&$\alpha  + \Omega  < 0.5,\ \displaystyle x_w > 1$ &${{\rho _\alpha }(x)}$\\ \hline
LD-MC& 
$\begin{array}{*{20}{c}}
{\beta  > 0.5,}\\
{0.5 - \Omega  < \alpha  < 0.5}
\end{array}$& ${\left\{ {\begin{array}{*{20}{l}}
{{\rho _\alpha }(x)}&{0 \le x \le {x_\alpha }}\\
{0.5}&{{x_\alpha } \le x \le 1}
\end{array}} \right.}$\\ \hline
LD-MC-HD& $\begin{array}{*{20}{c}}
{\alpha  < 0.5,\ \beta  < 0.5,}\\
{{x_\alpha } < {x_{\beta }}}
\end{array}$ & ${\left\{ {\begin{array}{*{20}{l}}
{{\rho _\alpha }(x)}&{0 \le x \le {x_\alpha }}\\
{0.5}&{{x_\alpha } \le x \le {x_\beta }}\\
{{\rho _\beta }(x)}&{{x_\beta } \le x \le 1}
\end{array}} \right.}$\\ \hline
MC-HD& $\begin{array}{*{20}{c}}
{\alpha  > 0.5,}\\
{0.5 - \Omega  < \beta  < 0.5}
\end{array}$ & ${\left\{ {\begin{array}{*{20}{l}}
{0.5}&{0 \le x \le {x_\beta }}\\
{{\rho _\beta }(x)}&{{x_\beta } \le x \le 1}
\end{array}} \right.}$\\ \hline
S& $\begin{array}{*{20}{c}}
{\alpha  < 0.5,\ \beta  < 0.5,}\\
{{x_\alpha } > {x_{\beta ,}}0 < x_w < 1}
\end{array}$ & ${\left\{ {\begin{array}{*{20}{l}}
{{\rho _\alpha }(x)}&{0 \le x \le {x_w}}\\
{{\rho _\beta }(x)}&{{x_w} \le x \le 1}
\end{array}} \right.}$\\ \hline
HD& ${\beta +\Omega < 0.5  ,\ x_w < 0}$ & $\rho_\beta(x)$\\ \hline 
MC& ${\alpha  > 0.5,\ \beta  > 0.5}$ & $0.5$\\ \hline \hline
\end{tabular}%

 \end{center}
\label{tab:K=1}
\end{table}

For reference, the particle steady-state densities and phase boundaries have been summarised in Table \ref{tab:K=1}. In the table, $\rho_{\alpha}(x)=\Omega x+\alpha,\ \rho_l(x)=0.5,$ and $\rho_{\beta}(x)=\Omega (x-1)+1-\beta$ corresponds to densities of low density phase (LD), maximal current phase (MC) and high density phase (HD), respectively.
The points separating the LD phase from the MC phase, and the MC phase from the HD phase are given by $x_\alpha=(1-2\alpha)/2\Omega$ 
and $x_\beta=(2\beta+2\Omega-1)/2\Omega $. Density discontinuity is located at the point $x_w=(\Omega-\alpha+\beta)/2\Omega$ in shock phase (S), where the currents for the left and the right solutions matches, $J_\alpha(x_w)=J_\beta(x_w)$.

\subsection{\textit{\textbf{\label{main} Unconserved ${V(m:n)}$ network of TASEP}}}
In the proposed network model $V(m:n)$, the absence of inter-segment interactions between ${L_k}'s$, forces all the $m$ incoming segments to behave identically, and thus they have the same phase among the seven possible phases (Table \ref{tab:K=1}). Similarly, all the $n$ outgoing segments have identical dynamics and hence all of them behave together having the same phase. Thus, the total number of phases cannot be greater than $7^2=49.$ 

We denote the particle current in any of the incoming segments ${L_k}'s$ ($k \in \{1,2,\cdots,$ $m\}$) by $J^{\text{in}}$ whereas in any outgoing segments ${R_k}'s$ ($k \in \{1,2,\cdots,n\}$), the current is denoted by $J^{\text{out}}$, respectively. It is assumed that the particles can leave any incoming segments ${L_k}'s$ from the last site with effective exit rate $\beta_{\text{eff}}$. Similarly, the effective entry rate of particles to the first site  of the ${R_k}'s$ segment is taken to be $\alpha_{eff}$. Now, utilising equation (\ref{bulklk}) which is obtained by performing the mean-field approximation, leads to the following non linear differential equation in the continuum limit for the average density profile ($\rho^{in}$),
\begin{equation}\label{bulklkk}
\frac{\partial\rho^{in}}{\partial t} +\frac{\partial J^{in}}{\partial x}=\Omega(1-2\rho^{in})
\end{equation}
along with the boundary conditions $\rho^{in}(0)=\alpha$ and $\rho^{in}(1)=1-\beta_{eff}$.
Here, $J^{in}$ denotes the average current in the incoming segments and is written as $\rho^{{in}}(1-\rho^{{in}})$. Similarly, the particle density $\rho^{out}$ in the outgoing segments satisfies
\begin{equation}\label{bulklk2}
\frac{\partial\rho^{out}}{\partial t} +\frac{\partial J^{out}}{\partial x}=\Omega(1-2\rho^{out})
\end{equation}
with the boundary densities, $\rho^{out}(0)=\alpha_{eff}$ and $\rho^{out}(1)=1-\beta$, respectively and the average current in the outgoing segments is given by $J^{out}=\rho^{out}(1-\rho^{out})$. \\
 Since the total current at the steady state is conserved throughout the system, we have 
\begin{equation}
 J=\sum\limits_{i = 1}^m {J^{in}}=\sum\limits_{i = 1}^n {J^{out}}\ \text{or}\ \  
  J=m {J^{in}}=n {J^{out}}
  \label{current}
\end{equation}
where $J$ denotes the overall current of the whole system.  Also, the current that leaves the left subsystem is equal to the current across the junction yields
\begin{equation}
m\beta_{eff}\rho^{in}(1)=m\rho^{in}(1)(1-\rho^{out}(0))
\label{out}.
\end{equation}
Similarly, the current across the junction must be equal to the current entering the right subsystem which gives
\begin{equation}
m\rho^{in}(1)(1-\rho^{out}(1))=n\alpha_{eff}(1-\rho^{out}(1)).
\label{in}
\end{equation}
Thus, from  equation (\ref{out}) and equation (\ref{in}), one can obtain the value of effective rates as
\begin{equation}
\alpha_{eff}=\frac{m}{n}\rho^{in}(1)\ \text{and}\ \beta_{eff}=1-\rho^{out}(0).
\label{effective}
\end{equation}
Utilising the current continuity condition given by equation (\ref{current}), we have
\begin{equation}
\mathop {\lim}\limits_{x \to {1^-}}
mJ^{in}=\mathop {\lim}\limits_{x \to {0^+}} nJ^{out}
\label{limit}.
\end{equation}

We adopt the effective entrance rate $\alpha_{eff}$ and exit rate $\beta_{eff}$ for the $L_k$ and $R_k$ segments along with equation (\ref{limit}) to investigate the possible structures of the system in terms of these rates. Our system has a large number of possible phases, so the dynamics of the whole system can be  understood by the phase transitions occurring across the junction. Thus, we categorize the phases into different subclasses based upon the nature of the phases near the junction. In each subclass, the methodology will be similar and hence the value of the effective rates, $\alpha_{eff}$ and $\beta_{eff}$ will remain the same.

Without loss of generality, we restrict our discussion to $m \geq n$. Depending upon how the topology changes near the junction, the phases can be divided into different categories. We designate the notation A $\rightarrow$ B to describe a subclass which denotes that the region just upstream to the junction in all the incoming segments is in A phase and the region just downstream to the junction of all the outgoing segments is in the B phase. 

Among the possible cases, there are certain cases which appear only for particular relation between the number of segments in the $L$ subsystem and the number of segments in the $R$ subsystem. For example, the cases MC $\rightarrow$ HD, MC $\rightarrow$ LD and MC $\rightarrow$ MC cannot exists when $m>n$. This is because if MC phase exists just upstream to the junction in the $L_k$ segments, then equation (\ref{current}) implies that $nJ^{out}=m/4$ which is not possible as $m>n$. Moreover, MC $\rightarrow$ LD junction do not even exists for $m=n$. In this case, the bulk densities around the junction are given by  $\rho^{{in}}=0.5$ and $\rho^{{out}}(x)=\Omega x+\alpha_{eff}$. Plugging these densities into equation  (\ref{limit}) gives $\alpha_{eff}=0.5$, which violates the existence condition of the phases in this subclass.

Now, for the case HD $\rightarrow$ LD, $\rho^{in}(1)=1-\beta_{eff}$ and $\rho^{out}(0)=\alpha_{eff}$ and using these in equation (\ref{effective}) yields $\alpha_{eff}=0$ and $ \beta_{eff}=1$, which violates the  conditions required for the existence of such phases when $m>n$ ($\beta_{eff}<1/2$). For $m=n$, equation (\ref{effective}) and equation (\ref{limit}) gives $\alpha_{eff}=\beta_{eff}=0.5$ which is again not possible. Analogous argument gives that LD $\rightarrow$ MC, MC $\rightarrow$ HD and HD $\rightarrow$ MC cannot exists for $m=n$. 
  We use the notation C:D to identify a phase in $V(m:n)$ network, where C and D describe a phase in all the incoming and the outgoing segments, respectively. We now discuss the stationary phases and existence conditions in each subclass, explicitly.
  
  \textbf{General case:} We analyze the subclasses which exists for all possible values of $m$ and $n$.
\begin{enumerate}
\item [(a)] \textbf{ LD $\rightarrow$ LD} - In this case, the region just upstream to the junction as well as the region just downstream to the junction display LD phase. Since the upstream region to the junction is in LD phase, so the entire incoming segment can exhibit only the low density phase. The possible choices for such phases are LD:LD, LD:LD-HD, LD:LD-MC and LD:LD-MC-HD. These phases are governed by the following common conditions,
\begin{equation}
\Omega+ \alpha<min\left\{\beta_{eff},\displaystyle\frac{1}{2}\right\},\quad \alpha_{eff}<\displaystyle\frac{1}{2}.
\label{LDconditions}
\end{equation}
Here, the bulk density in each of the incoming segments is
$\rho^{in}(x)=\Omega x+\alpha $ whereas the bulk density in the outgoing segments near to the junction is $\rho^{out}(x)=\Omega x+\alpha_{eff}$. 
To determine the explicit particle densities in such phases, the effective rates can be calculated by utilizing the above expressions of density and equation \eqref{LDconditions} in equation \eqref{limit}, that yields, 
\begin{equation}
\alpha_{eff}=\displaystyle\begin{cases}
 \frac{1}{2}\left(1-\sqrt{1-\frac{4m(\Omega+\alpha)(1-\Omega-\alpha)}{n}}\right) & m>n\\
\Omega+\alpha & m=n.
\end{cases}
\label{LDeff} 
\end{equation}
This equation holds only when $\Omega+\alpha\leq\gamma$ where \\
\begin{equation}
 \gamma= \begin{cases} 
    \displaystyle\frac{1}{2}\left(1-\sqrt{1-\frac{n}{m}}\right) & m>n \\
    \displaystyle\frac{1}{2} & m=n.
   \end{cases}
\end{equation}
The density at the first site of the $R_k$ segments is
%Since chain III is in low density phase near the junction, 
$\displaystyle\rho^{out}(0)=\alpha_{eff}$ and by equation\eqref{effective}, $\displaystyle\beta_{eff}=(1-\alpha_{eff})$. Table \ref{tab:LD-B phase} shows the various phases in this case along with the conditions for the existence of the corresponding dynamic regime. 
\begin{table}[htb] 
\begin{center}
\caption{Phases with LD to LD (LD $\rightarrow$ LD) transition at the junction.}

\begin{tabular}{||c|c||} 

 \hline \hline
Phase & Phase Boundary   \\ 
 \hline \hline
 LD:LD & $\Omega+ \alpha_{eff}<\text{min}\{\beta,0.5\}$   \\     \hline 
 LD:S &  $\beta-\Omega<\alpha_{eff}<\text{min}\{\beta+\Omega,\ 1-\beta-\Omega$\} \\   \hline
 LD:LD-MC & $\Omega  + \alpha  < \gamma,\ 0.5<\Omega  + \alpha_{eff},\ 0.5<\beta$\\ \hline
LD:LD-MC-HD & $\Omega  + \alpha  < \gamma,\ 1-\Omega -{\alpha _{eff}}<\beta  < 0.5$  \\ 

 \hline \hline
\end{tabular}
\label{tab:LD-B phase}
\end{center}\end{table}
\item[(b)] \textbf{LD $\rightarrow$ HD} - The only possible phase in this subclass is LD:HD. This phase occurs when all the $L_k$ segments are in LD phase and all the $R_k$ segments are in HD phase. Such phase is specified by the conditions,
\begin{equation}
\Omega+\alpha<\text{min}\left\{\beta_{eff},\frac{1}{2}\right\},\quad \beta+\Omega<\text{min}\left\{\alpha_{eff},\frac{1}{2}\right\}
\label{LD:HDc}.
\end{equation}
The corresponding equations for bulk densities are $\rho^{in}(x)=\Omega x+\alpha$ and $\rho^{out}(x)=\Omega (x-1)+1-\beta.$
Utilising equation \eqref{limit}, we obtain
\begin{equation}
\beta= \begin{cases}
\displaystyle
 \frac{1}{2}\left(1-\sqrt{1-\frac{4m(\Omega+\alpha)(1-\Omega-\alpha)}{n}}\right)-\Omega & m>n\\
\alpha & m=n.
\end{cases}
\label{LD:HDeff} 
\end{equation} 
The above equation is valid only when 
\begin{equation}
\Omega+\alpha\leq \begin{cases}
\displaystyle\frac{1}{2}\left(1-\sqrt{1-\frac{n}{m}}\right) & m>n\\
\displaystyle\frac{1}{2} & m=n.
\end{cases}
\label{LD:HD}
\end{equation}

\item [(c)] \textbf{HD $\rightarrow $ HD} - Here, all the segments  $L_k$ as well as $R_k$ portray HD phase. The phases which fall under this category are HD:HD, MC-HD:HD, LD-MC-HD:HD and LD-HD:HD phase. The HD phase in the outgoing segments can exists when 
\begin{equation}
\Omega+ \beta<min\left\{\alpha_{eff},\displaystyle\frac{1}{2}\right\}.
\label{HDc}
\end{equation}
By equation \eqref{limit}, the boundary parameters must satisfy 
\begin{equation}
\beta_{eff}=\begin{cases}
\frac{1}{2}\left(1-\sqrt{1-\frac{4n(\Omega+\beta)(1-\Omega-\beta)}{m}}\right) &m>n\\
\Omega+\beta&m=n
\end{cases}
\label{HDeff} 
\end{equation}
Moreover, $\rho^{in}(1)=1-\beta_{eff}$ and  utilising equation  \eqref{effective} to obtain the value of $\alpha_{eff}=\displaystyle\frac{m}{n}(1-\beta_{eff})$.
All the desirable phases of this case with the parameter ranges are summarised in Table \ref{tab:A-HD phase}.

\begin{table}[h]
\begin{center}
\caption{Phases with HD to HD (HD $\rightarrow$ HD) transition at the junction.}

\begin{tabular}{||c|c|c||} 

 \hline \hline
 Phase & \multicolumn{2}{|c||}{Phase Boundary }  \\ \cline{2-3}
 & $m> n$ & $m=n$ \\
 \hline \hline
 LD-MC-HD:HD & \multicolumn{2}{|c||}{$1-\beta_{eff}-\Omega<\alpha<0.5,\ \beta<0.5-\Omega$} \\ \hline  
 S:HD & \multicolumn{2}{|c||}{$\alpha-\Omega< \beta_{eff}< \Omega+\alpha$} \\    \cline{2-3}
 &  $ \beta<0.5-\Omega$ &$\beta_{eff}<1-\Omega-\alpha$ \\   \hline
MC-HD:HD& $0.5<\alpha$ &$0.5<\alpha$ \\
& $\beta  < 0.5-\Omega$ &$\beta  < 0.5-\Omega<\beta_{eff}$ \\ \hline
HD:HD & $\Omega+\beta_{eff}<\alpha$ & $\Omega+\beta_{eff}<\text{min}\{\alpha,0.5\}$  \\ 
 & $\Omega+\beta<0.5$ &   \\ 

 \hline \hline
\end{tabular}
\label{tab:A-HD phase}
\end{center}
\end{table}
\end{enumerate}
\textit{\textbf{Special cases:}} Now we will discuss the possibility of the existence of the phases which exist for specific relation between the number of incomings and the number of outgoing segments. The subclasses LD $\rightarrow$ MC and HD $\rightarrow$ MC exists only when $m>n$ whereas the subclass MC $\rightarrow$ MC prevail only for $m=n$. We explore the existence conditions and phase boundaries for these subclasses. 

\begin{enumerate}
\item[(a)]  \textbf{LD $\rightarrow$ MC} - As discussed earlier this  case only exists when $m>n$. We assume that all the $L_k$ segments are in low density phase whereas the $R_k$ segments show maximum current  phase at the upstream boundary provided the boundary parameters satisfy the following common relations, 
\begin{equation}
\Omega  + \alpha  < \text{min}\left\{{\beta _{eff},\ \frac{1}{2}}\right\},\ {\alpha _{eff}} > \frac{1}{2}
\label{LD:MCc}
\end{equation}
\begin{table}[h] 
\caption{\label{LD to MC phase} Phases with LD to MC (LD $\rightarrow$ MC) transition at the junction for $m>n.$}
\begin{center}

\begin{tabular}{||c| c||} 

 \hline \hline
 Phase & {Phase Boundary }  \\ \hline
 \hline
 LD:MC &   $\Omega  + \alpha  =\displaystyle\frac{1}{2}\left(1-\sqrt{1-\frac{n}{m}}\right),\ \beta>\displaystyle\frac{1}{2}$ \\ \hline
  LD:MC-HD   &$\Omega  + \alpha  = \displaystyle\displaystyle\frac{1}{2}\left(1-\sqrt{1-\frac{n}{m}}\right),\ \displaystyle\frac{1}{2}-\Omega<\beta  < \displaystyle\frac{1}{2}$ \\
  \hline \hline
 
\end{tabular}
 \end{center}
\end{table}

The feasible choices for phases, in this case, are LD:MC and LD:MC-HD.
The particle densities near the junction are given by
$
\rho^{in}(x)=\Omega x+\alpha$ and $\rho^{out}(x)=0.5$. By current continuity condition equation \eqref{limit}, we get
\begin{equation}
\Omega  + \alpha  =
 \displaystyle\frac{1}{2}\left(1-\sqrt{1-\frac{n}{m}}\right).
\label{LD:MCeff} 
\end{equation}
The various phases in this case for $m>n$ are shown in Table \ref{LD to MC phase} along with the conditions for the existence of the corresponding dynamic regime.

\item[(b)] \textbf{HD $\rightarrow$ MC} - As mentioned earlier, this subclass do not exists for $m=n$. Here, the downstream part of all the $L_k$ segments are exit dominated and the upstream portion of the $R_k$ segments manifest
\begin{table}[h]
\begin{center}
\caption{\label{tab:HD-MC phase}Phases with HD to MC (HD $\rightarrow$ MC) transition at the junction for $m>n.$}

\begin{tabular}{||c| c||} 

 \hline \hline
 Phase & Phase Boundary  \\ 
 \hline \hline
HD:MC &   $\Omega  + {\beta _{eff}} < \alpha ,\ \beta  > 0.5$ \\ \hline
S:MC   &$\alpha-\Omega  <{\beta _{eff}} < \Omega  + \alpha,\ \beta  > 0.5 $ \\ \hline
  S:MC-HD &   $\alpha-\Omega < {\beta _{eff}}< \Omega  + \alpha$\\ 
&   $0.5 - \Omega  < \beta  < 0.5$\\ \hline
 HD:MC-HD & $  \Omega  + {\beta _{eff}} < \alpha,\  0.5 - \Omega  < \beta  < 0.5$ \\ \hline
 MC-HD:MC-HD & $0.5-\Omega<\beta<0.5,\ \alpha<0.5$  \\ \hline
  MC-HD:MC & $\beta>0.5,\ \alpha>0.5$   \\ \hline
 LD-MC-HD:MC &  $1-\beta_{eff}-\Omega<\alpha<0.5,\ \beta>0.5$  \\ \hline
 LD-MC-HD:MC-HD &  $1-\beta_{eff}-\Omega<\alpha<0.5$  \\
 &  $ 0.5-\Omega<\beta<0.5$  \\
 \hline \hline
\end{tabular}
\end{center}

\end{table}
maximal current. So, the possible phases in this category are HD:MC, LD-HD:MC, MC-HD:MC, LD-MC-HD:MC, LD-HD:MC-HD, HD:MC-HD, MC-HD:MC-HD and LD-MC-HD:MC-HD. 
The density of particle in the bulk around the junction in each incoming segment is $\rho^{in}(x)=\Omega(x-1)+1-\beta$ and in that of outgoing segments is $\rho^{out}(x)=0.5$. The effective rates $\alpha_{eff}$ and $\beta_{eff}$ can be determined by substituting the above values of particle densities in equation \eqref{limit} to  obtain
\begin{equation}
\beta_{eff}=
\displaystyle \frac{1}{2}\left(1-\sqrt{1-\frac{n}{m}}\right).
\label{betaeff}
\end{equation} 
Furthermore,  in this case, $\rho^{in}(1)=1-\beta_{eff}$ and by equation \eqref{effective},  $\alpha_{eff}=\displaystyle \frac{m}{n}\left(1-\beta_{eff}\right)$. Summary of all the desirable phases of this case with the parameter ranges for $m>n$ is given in Table \ref{tab:HD-MC phase}.

\item[(c)] \textbf{MC $\rightarrow$ MC} - As maximal current phase can exists both upstream and downstream to the junction only when $m=n$, so MC $\rightarrow$ MC phase transition occurs only for equal number of incoming and outgoing segments. In this phase, the downstream boundaries of all the ${L_k}'s$ and the upstream boundaries of all the ${R_k}'s$ are in maximal current phase. The probable phases here can be LD-MC:MC, MC:MC, LD-MC:MC-HD and MC:MC-HD.
 Table \ref{MC to Mc} summarises all the desirable phases of this case with the parameter ranges for $m=n$.
\begin{table}[h!]
\begin{center}
\caption{Phases with MC to MC (MC
 $\rightarrow$ MC) transition at the  junction.}
\begin{tabular}{||c| c||} 
 \hline \hline
 Phase & Phase Boundary  \\ 
 \hline \hline
 LD-MC:MC &   $0.5-\Omega<\alpha  < 0.5,\ \beta>0.5$ \\ \hline
 MC:MC &   $\alpha  > 0.5,\ \beta>0.5$ \\ \hline
 LD-MC:MC-HD &   $0.5-\Omega<\alpha  < 0.5,\ 0.5-\Omega<\beta<0.5$ \\ \hline
 MC:MC-HD &   $\alpha  > 0.5,\ 0.5-\Omega<\beta<0.5$ \\   
 \hline \hline
\end{tabular}
\label{MC to Mc}
\end{center}
\end{table}

\end{enumerate}

The theoretical observations based upon the mean field argument predict that out of the $49$ possible phases, $30$ phases are not realized in the system and for the remaining $19$ phases the existing conditions have been thoroughly discussed above for $m>n$.When the number of segments in both subsystems are equal i.e., $m=n$, the number of admissible phases reduced to $13$ and the remaining $36$ phases cease to exists.
For the limiting case $\Omega=0$, the system exhibits only five phases namely, LD:LD, LD:HD, LD:MC, HD:MC and HD:HD, out of the $19$ potential phases described above. These findings match very well with reference \cite{wang2008theoretical,zhang2019exclusion} which are particular scenarios of our model for $n=1$, hence depicting the accuracy of our theoretical results.

 \section{\label{Results Discussions}Results \& Discussions}
In this section, we exploit the general conditions of existence discussed in the previous section to address the behaviour of the system in the $\alpha-\beta$ plane. Our main aim is to explore the effect of the LK rates on the stationary properties of the system. Furthermore, we intend to investigate the effect of the number of segments in each subsystem on the topology of the phase diagram.  The theoretical outcomes have been extensively validated through Monte Carlo simulations. We have adopted the random sequential updating rule and the number of sites in each segment is considered to be $N = 1000$. To ensure the occurrence of steady state, first $5\%$ of the time steps are discarded and the average density of particles is computed over an interval of $10N$. The phase boundaries are computed
within an estimated error of less than $2\%$. For the thorough discussion, we analyze the system dynamics for two different categories: (i) when the number of segments in both subsystems are equal $(m=n)$ and (ii) when the number of segments in both subsystems are different $(m\neq n)$. 
\subsection{\label{mequaln}
\textit{\textbf{$m=n$}}}

\begin{figure*}[!htb]
	\centering
	\subfigure[\label{fig:Omega=0n}]{\includegraphics[width = 0.45\textwidth]{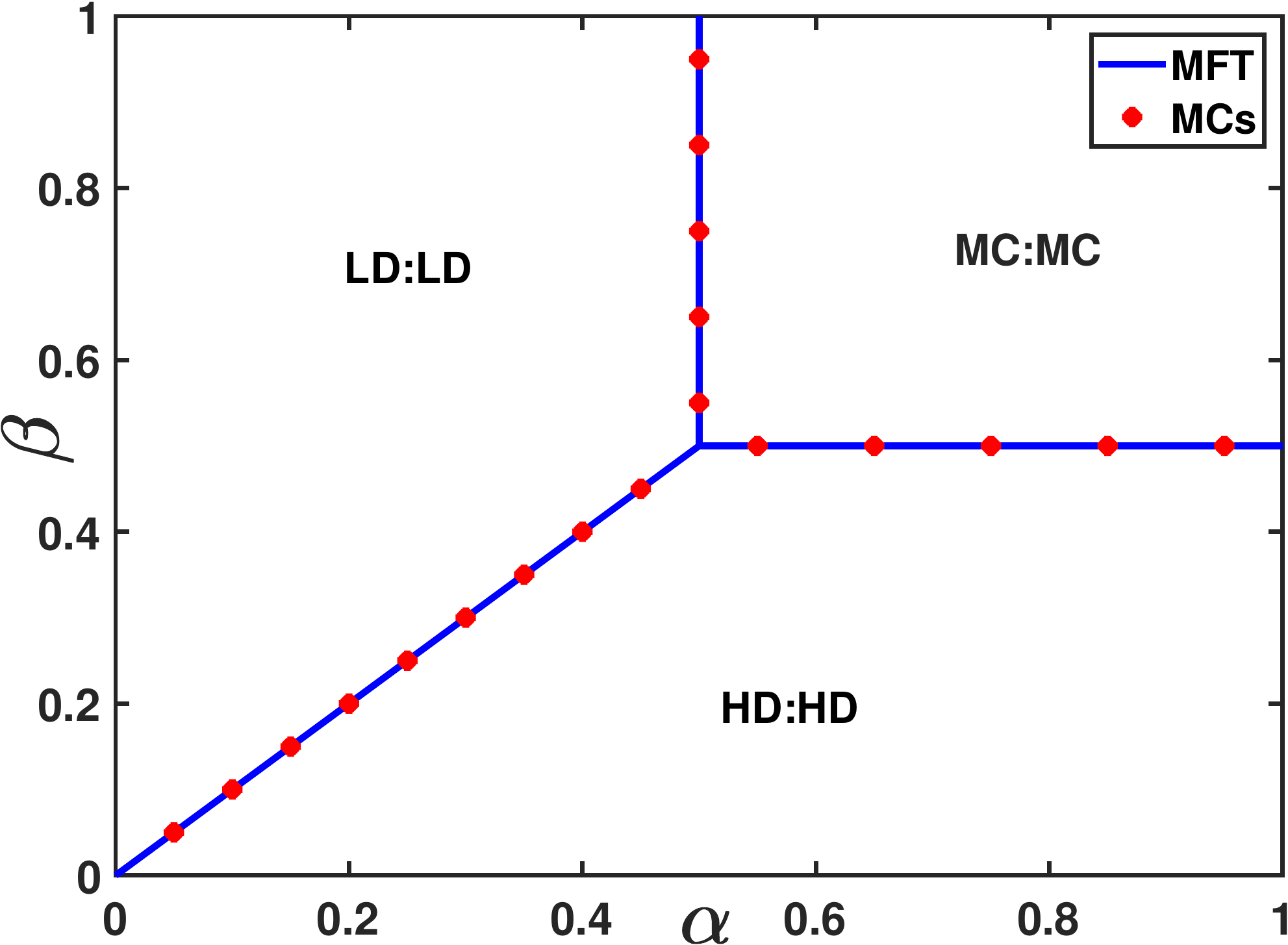}} \quad \quad
	\subfigure[\label{fig:Omega=0.1n}]{\includegraphics[width = 0.45\textwidth]{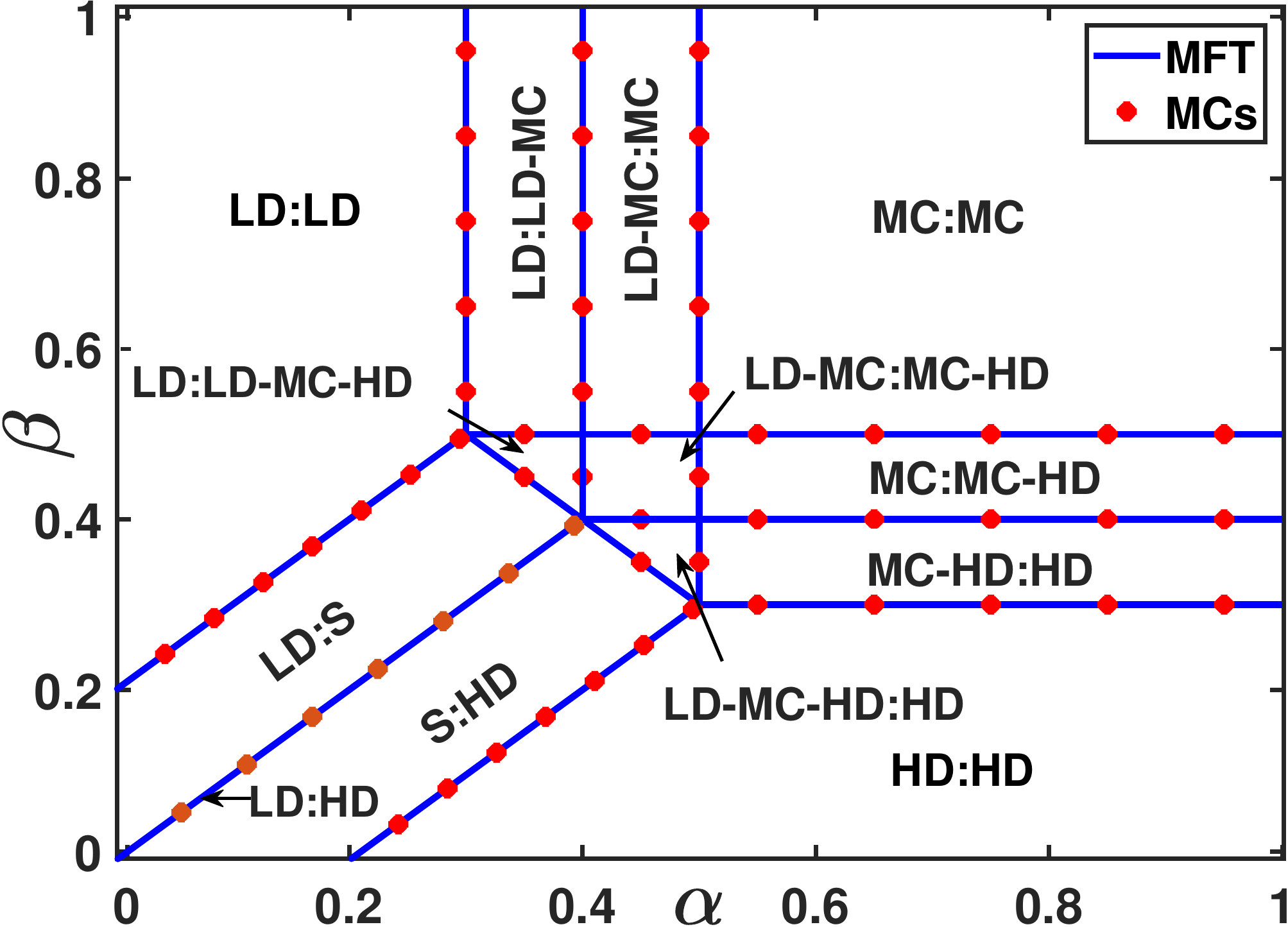}}\\
    \subfigure[\label{fig:Omega=0.25n}]{\includegraphics[width = 0.45\textwidth]{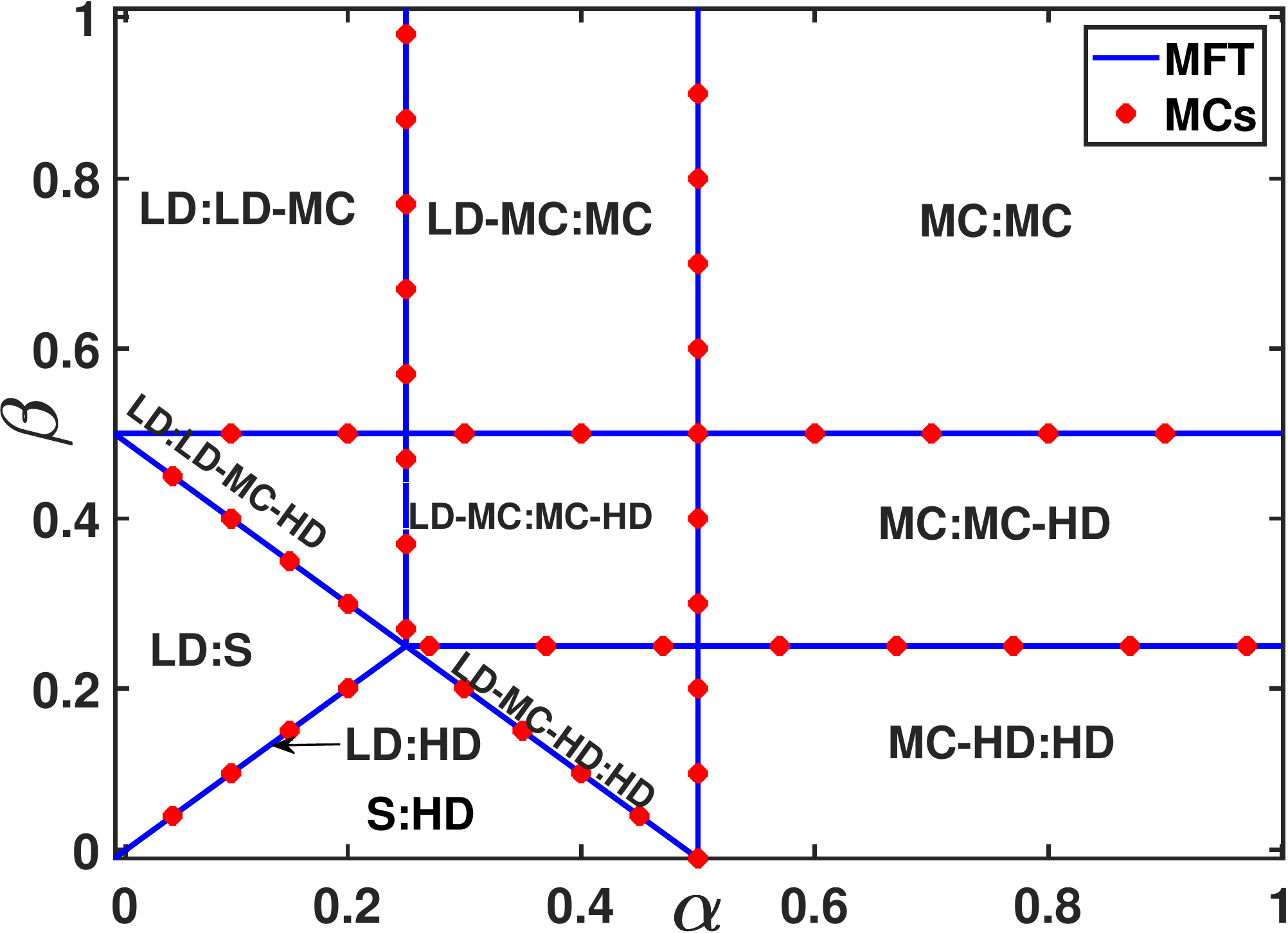}}\quad \quad
	\subfigure[\label{fig:Omega=0.5n}]{\includegraphics[width = 0.45\textwidth]{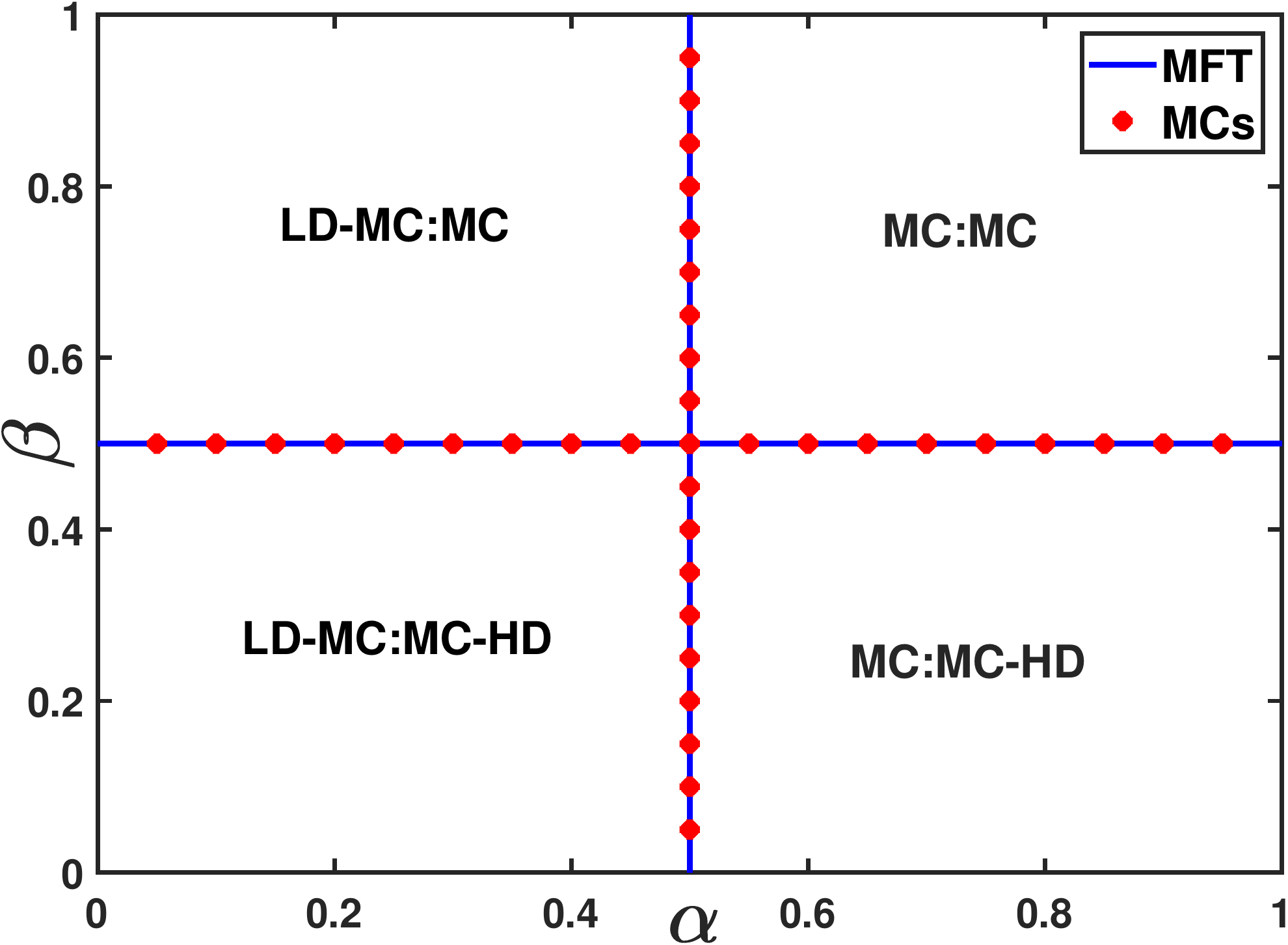}}
	\caption{\label{phase}Stationary phase diagrams for different value of $\Omega$: (a) $\Omega=0$, (b) $\Omega=0.1$, (c) $\Omega=0.25$ and (d) $\Omega=0.5$  for $m=n=2$ . Note that, the phase diagrams as well as the density profiles will remain unaffected for any values of $m$, $(n=m)$. The phase transformations are continuous for boundaries between all observed phases. Solid blue lines represents theoretical results and dotted red symbols correspond to Monte Carlo simulation (MCs).}
\end{figure*}

We begin our analysis for the case when the number of segments in each subsystem are equal and investigate the non-trivial effects on the topology of the phase diagram with the governing parameter $\Omega$ in the $\alpha-\beta$ plane. For the inspection, we specifically consider $m=n=2$ which can be generalized for any values of $m$ and $n$. For the limiting case $\Omega=0$, the phase diagram consists of only three phases viz., LD:LD, MC:MC and HD:HD which resembles with the phase diagram for single-segment TASEP as illustrated in figure \ref{fig:Omega=0n}.  As soon as LK dynamics is introduced in the system, the number of feasible phases increase drastically with the emergence of thirteen stationary phases namely, LD:LD, LD:LD-MC-HD, LD:S, LD:HD, S:HD, HD:HD, LD:LD-MC, LD-MC:MC, LD-MC:MC-HD, MC:MC, MC:MC-HD, MC-HD:HD, and LD-MC-HD:HD as shown in figure \ref{fig:Omega=0.1n} for $\Omega=0.1$. It is interesting to note that among these phases, the system can sustain maximal current in both the upstream and downstream of the junction as the number of segments in both the $L$ and $R$ subsystems are equal. With an increase in $\Omega$, only the re-positioning of the phase boundaries takes place till a critical value $\Omega_{c_1}=0.25$. This value of $\Omega_{c_1}$ can be theoretically  obtained from equation \eqref{LDconditions} and after this value LD:LD and HD:HD no longer persists in the system as shown in figure \ref{fig:Omega=0.25n} for $\Omega=0.25$. The remaining eleven phases continue to persists until a critical value $\Omega_{C_2}=0.5$, as evident from equation \eqref{LD:MCeff}. Beyond this critical value, the topology of the phase diagram becomes most simplified where only four phases are realised namely, LD-MC:MC, MC:MC, LD-MC:MC-HD and MC:MC-HD as illustrated in figure \ref{fig:Omega=0.5n} for $\Omega=0.5$. Now, as $\Omega \rightarrow \infty$, LK  rates dominate the overall dynamics of the system and only these four phases continue to exists.

\begin{figure}[htb]
\begin{center}
\includegraphics[width=8cm, height=6cm]{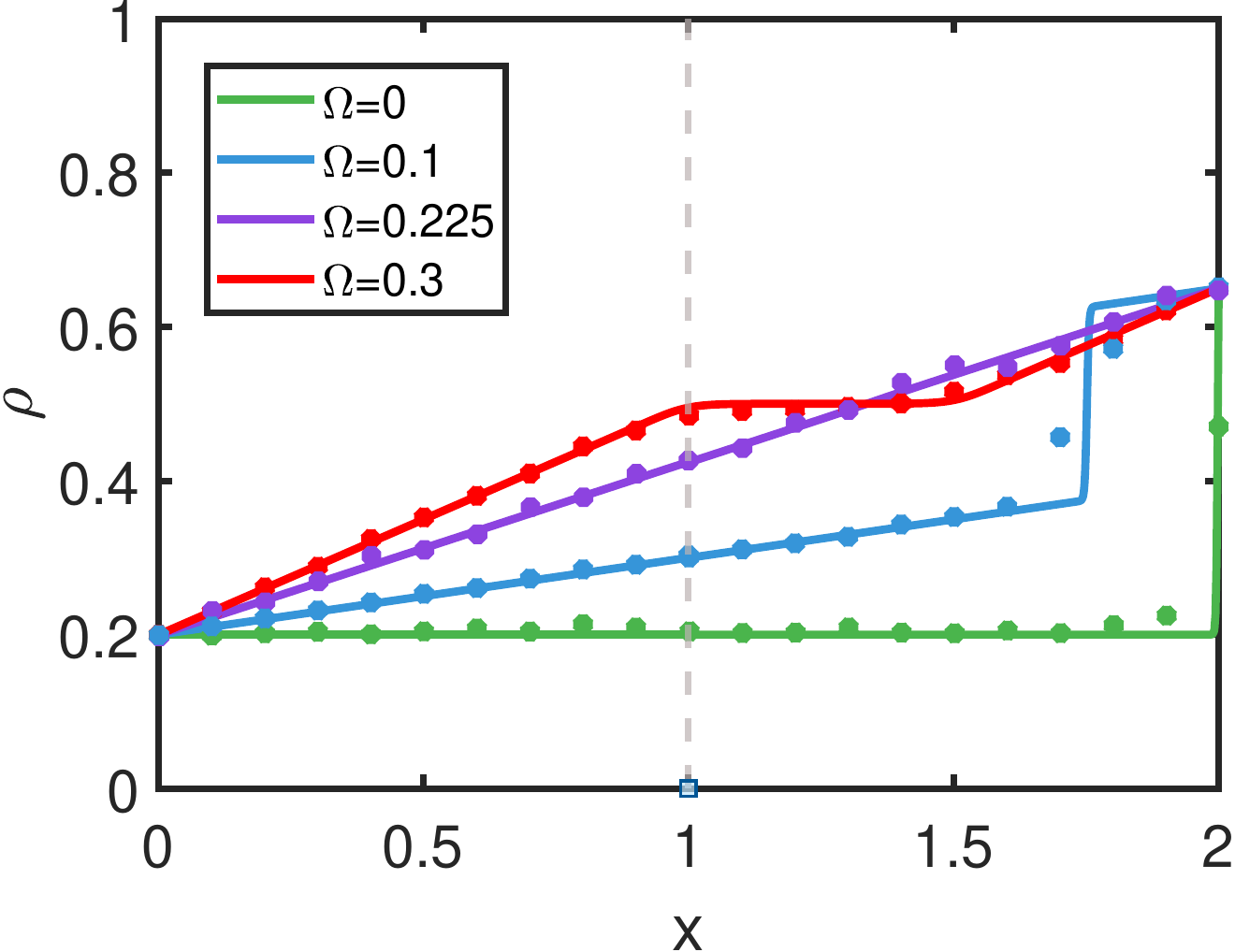}
\caption{Density profiles for different value of $\Omega$ when $m=n=2$. The vertical line at $x=1$ shows the position of the junction.}
\label{transition}
\end{center}
\end{figure}

To understand how the phase transitions occur with an increase in $\Omega$, we have plotted the density profiles keeping fixed the boundary controlling parameters $\alpha=0.2$ and $\beta=0.35$ (see figure \ref{transition}). For these values of $\alpha$ and $\beta$, the system exhibits LD:LD phase for $\Omega=0$. With an increase in $\Omega$, the density profile transits from LD:LD to LD:S then to LD:LD-MC-HD and finally to LD-MC:MC-HD. This can be explained by the following arguments. With an increase in $\Omega$, the value of $\alpha_{eff}$ increases whereas $\beta_{eff}$ decreases. The system is governed by the LK dynamics leading to the accumulation of particles in the outgoing segments and hence LD:S phase is observed. At $\Omega=0.225$, the density in the outgoing segments becomes a 
simple linear profile, continuously matching the density induced by $\alpha_{eff}$ and $\beta$. Consequently, the shock phase vanishes and LD-MC-HD phase emerges in the outgoing segments. As $\Omega$ is increased until $0.3$, $\alpha_{eff}\to 0.5^-$ and $\beta_{eff}\to 0.5^+$, thus maximal current is observed around the junction and the system manifest LD-MC:MC-HD phase.

Though we have provided the results for $m=n=2$ but the findings will remain unaffected for any values of $m$ and $n$ $(m=n)$. This is because of the reason that all the phase boundaries are independent of the number of incoming segments and the outgoing segments. Attributable to this fact the topology of the phase diagram obtained for $V(m:m)$ network closely resembles  that of a single-segment TASEP-LK model. This can be easily verified by considering the following phase boundary transformations:\\
\begin{align*}
 \text{$V(m:m)$ network} &\leftrightarrow \text{Single segment TASEP-LK}\\
 \Omega+\alpha_{eff}=\text{min}\{\beta,0.5\}&\leftrightarrow \Omega+\alpha=\text{min}\{\beta,0.5\}\\
 \alpha=0.5&\leftrightarrow\alpha=0.5\\
 2\Omega+\beta+\alpha=1 &\leftrightarrow  \Omega+\beta+\alpha=1\\
 \Omega+\beta_{eff}=\text{min}\{\alpha,0.5\}&\leftrightarrow \Omega+\beta=\text{min}\{\alpha,0.5\}\\
\beta=0.5&\leftrightarrow\beta=0.5\\
  \end{align*}
whereas for the phases\\
\begin{table}[h]
\begin{center}

\begin{tabular}{rcl}
LD:LD~~~~ & $\leftrightarrow$ &LD\\ 
\smallskip
$\left. {\begin{array}{*{20}{r}}
{\text{LD:LD-MC}}\\
{\text{LD-MC:MC}}
\end{array}} \right\}$& $\leftrightarrow$& LD-MC\\
\smallskip
$\left. {\begin{array}{*{20}{r}}
{\text{MC:MC-HD}}\\
{\text{MC-HD:HD}}
\end{array}} \right\}$& $\leftrightarrow$& MC-HD\\ 
\smallskip
$\left. {\begin{array}{*{20}{r}}
{\text{LD:S}}\\
{\text{S:HD}}\\
{\text{LD:HD}}
\end{array}} \right\}$& $\leftrightarrow$& LD-HD\\
$\left. {\begin{array}{*{20}{r}}
{\text{LD:LD-MC-HD}}\\
{\text{LD-MC:MC-HD}}\\
{\text{LD-MC-HD:HD}}
\end{array}} \right\}$& $\leftrightarrow$& LD-MC-HD\\
HD:HD~~~~&$\leftrightarrow$ &HD\\
MC:MC~~~~&$\leftrightarrow$ &MC.\\
\end{tabular}
\end{center}
\end{table}
\subsection{\textbf{$m\neq n$}}

\begin{figure*}[!htb]
	\centering
	\subfigure[\label{fig:Omega=0}]{\includegraphics[width = 0.32\textwidth]{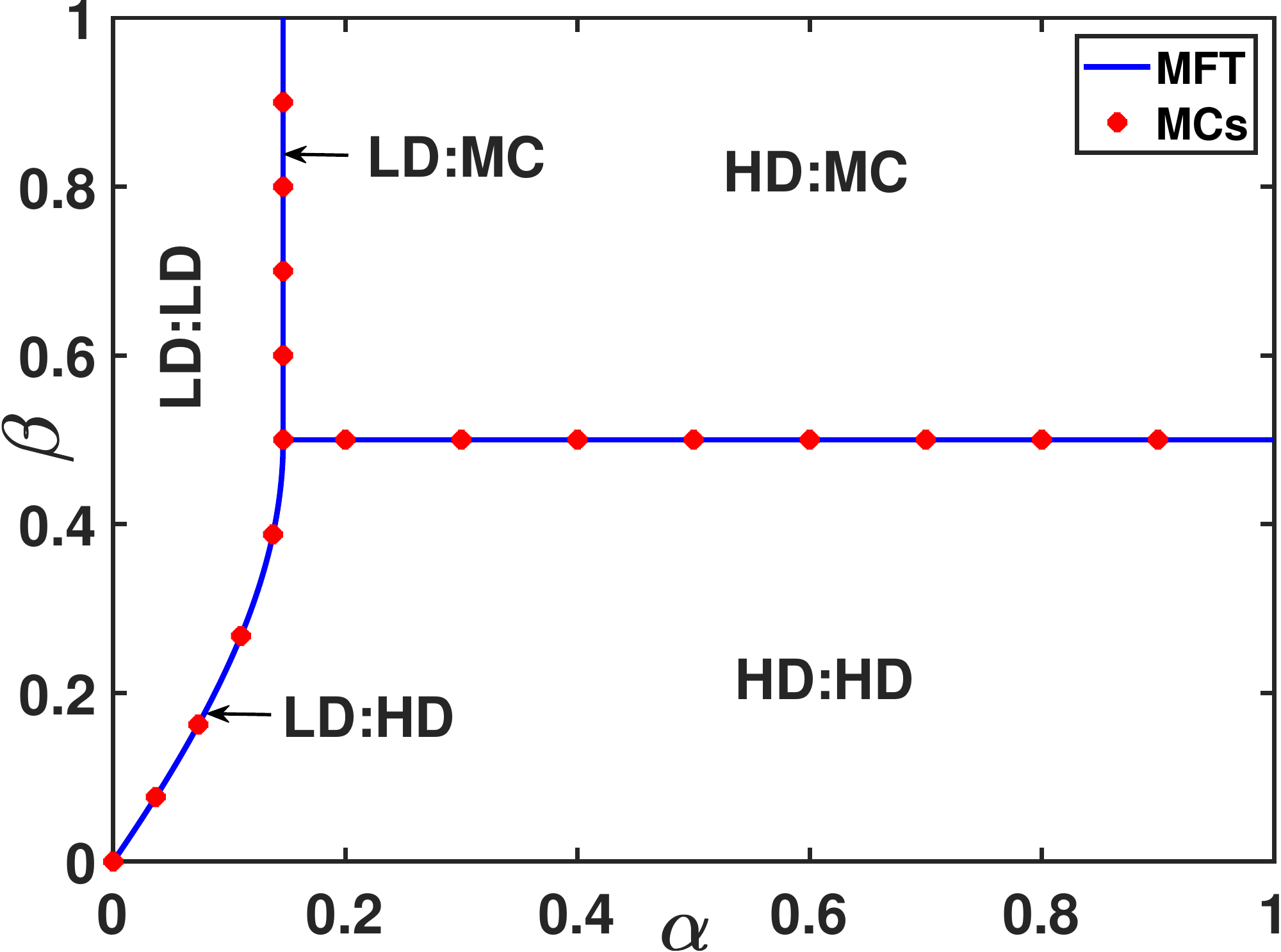}} \quad \quad
	\subfigure[\label{fig:Omega=0.1new}]{\includegraphics[width = 0.32\textwidth]{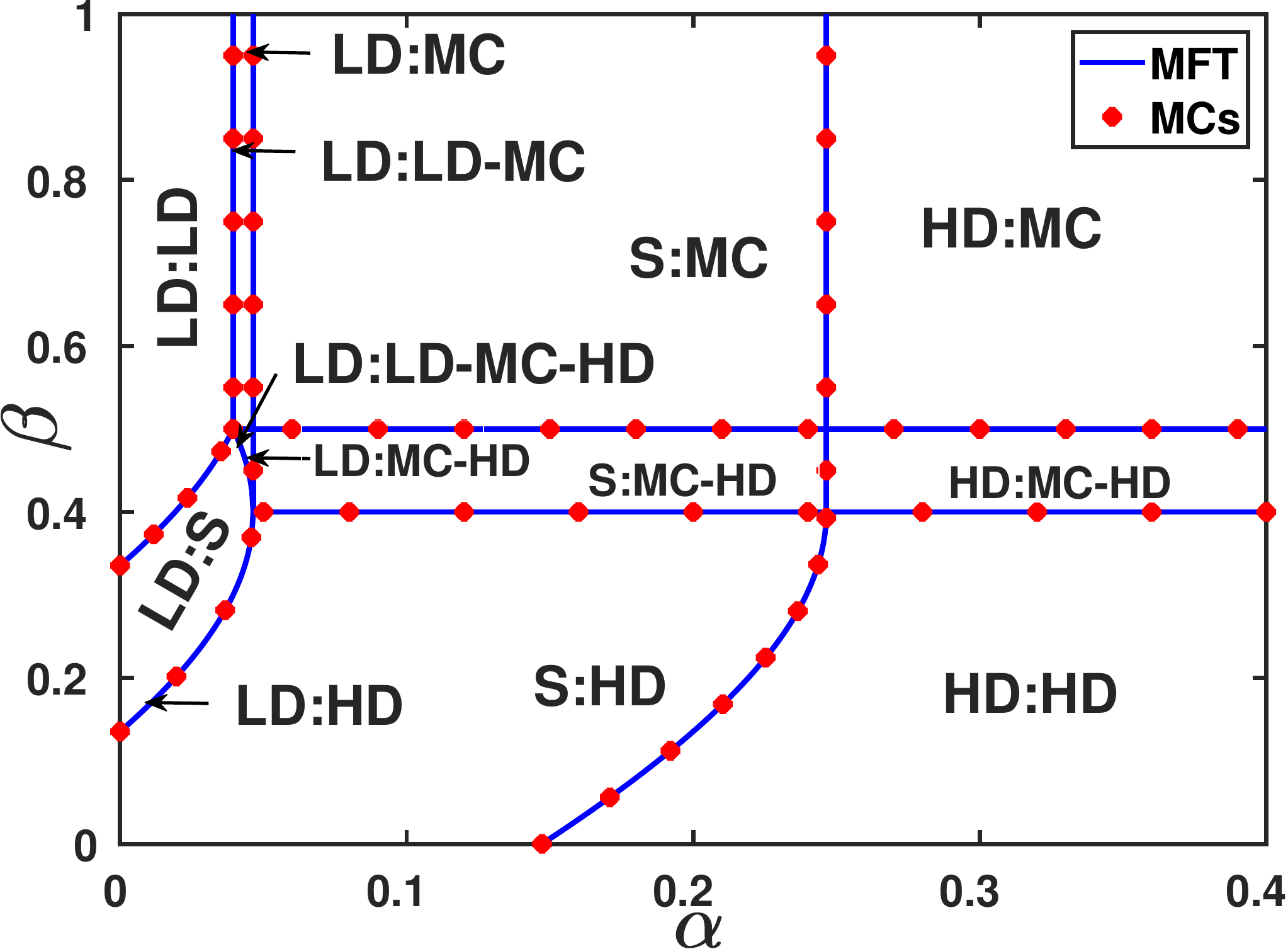}}\\
    \subfigure[\label{fig:Omega=0.133}]{\includegraphics[width = 0.32\textwidth]{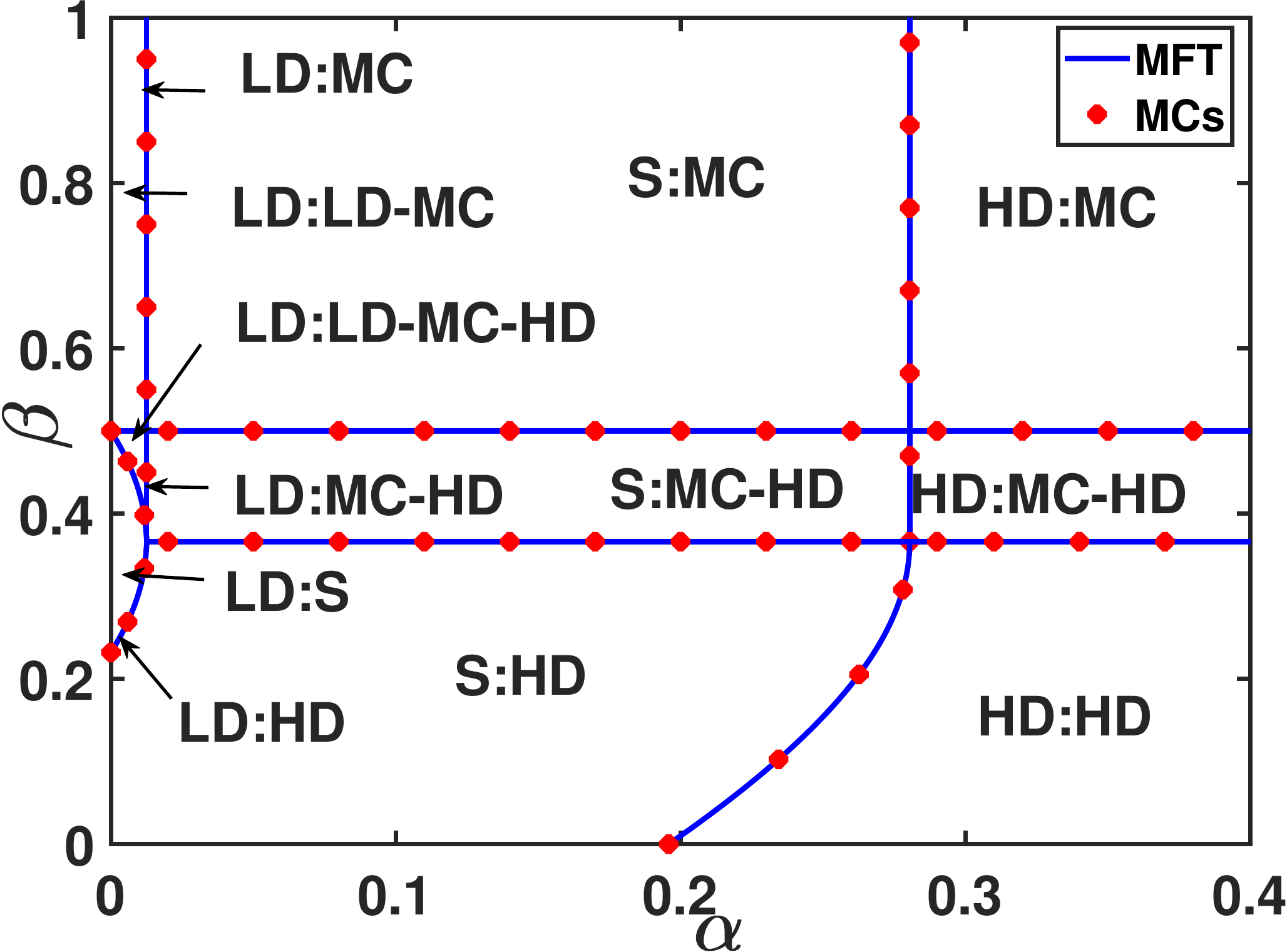}}\quad \quad
	\subfigure[\label{fig:Omega=0.1464new}]{\includegraphics[width = 0.32\textwidth]{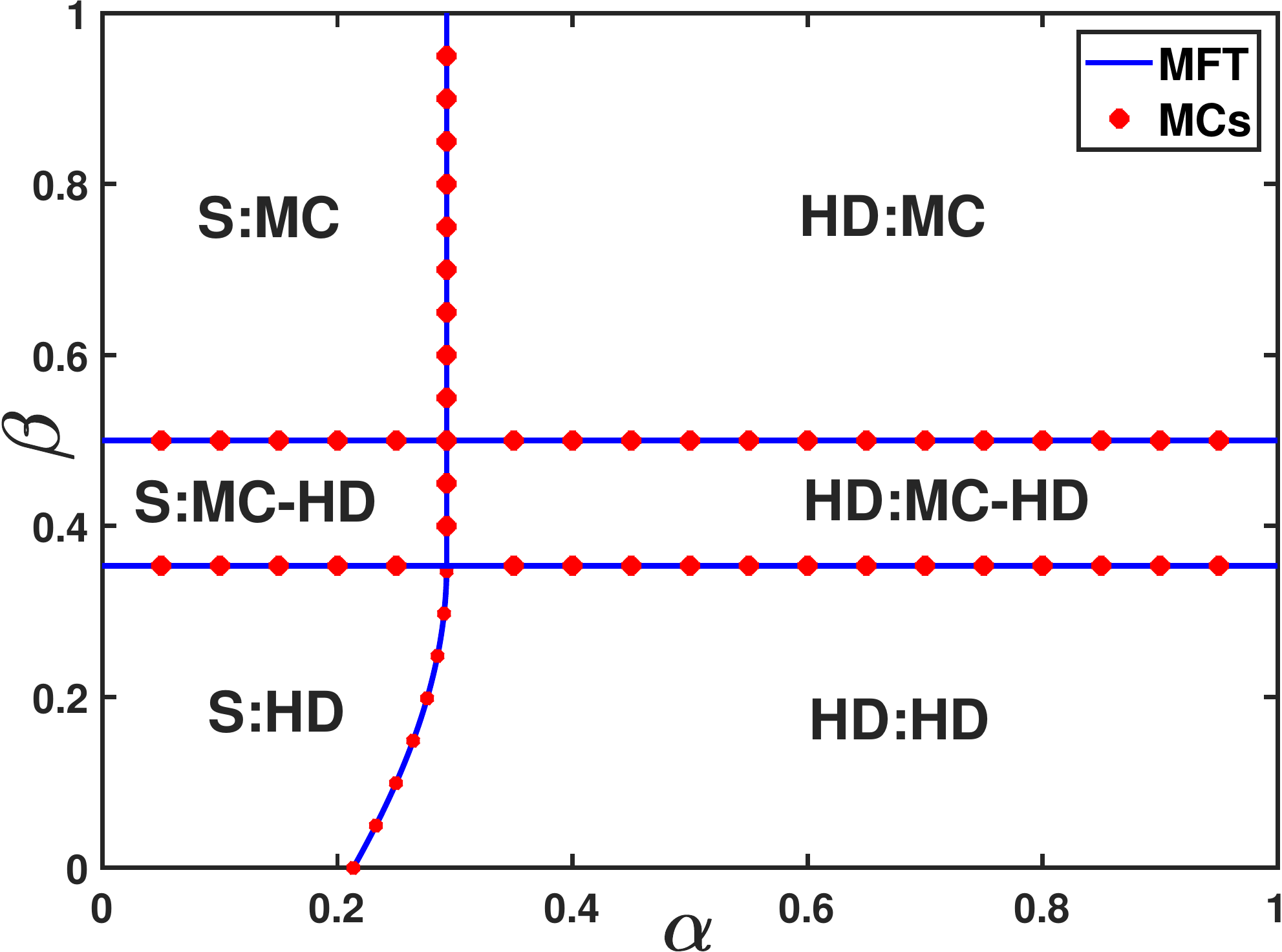}}\\
	\subfigure[\label{fig:Omega=0.356}]{\includegraphics[width = 0.32\textwidth]{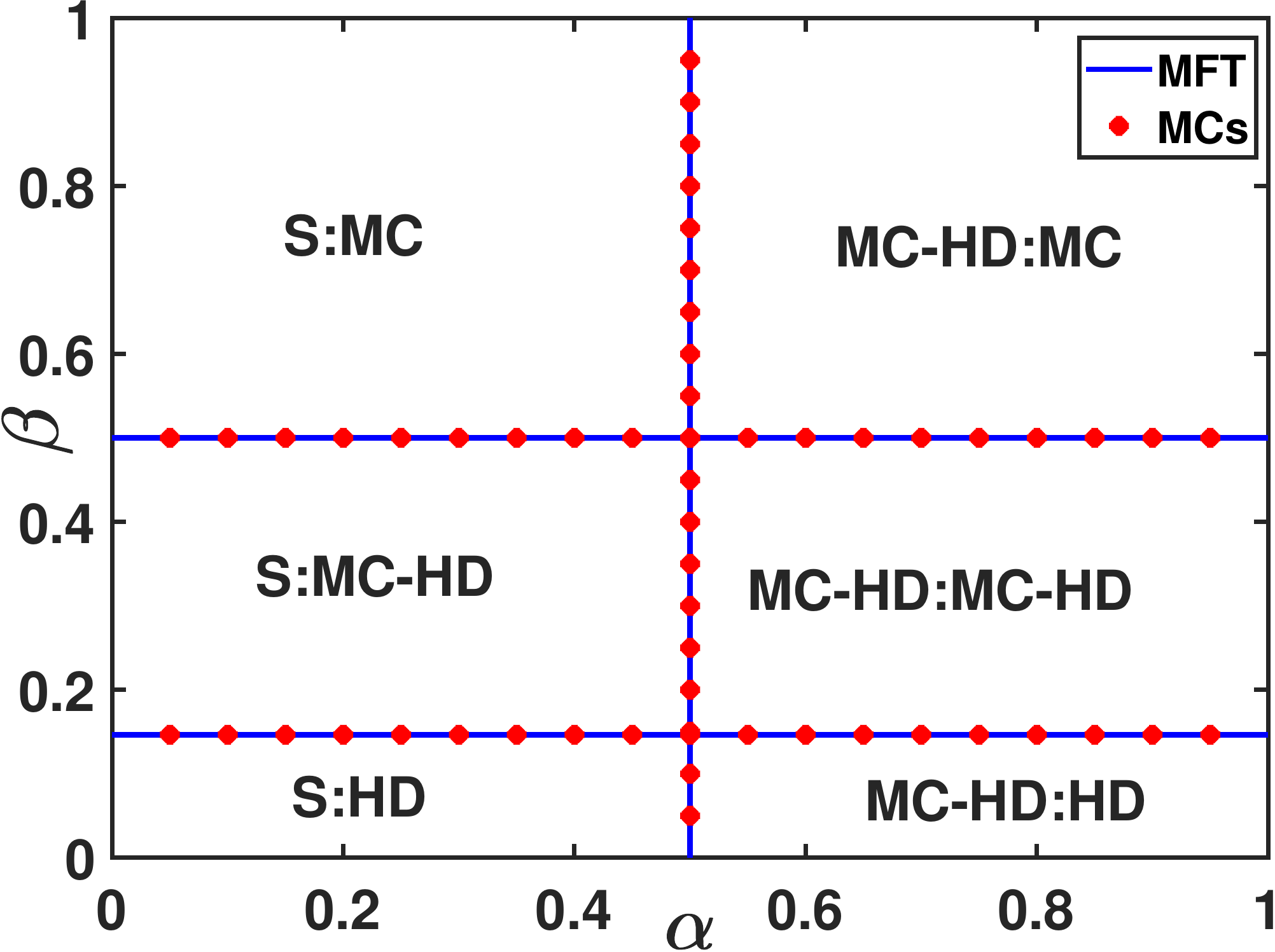}}\quad \quad
	\subfigure[\label{fig:Omega=0.42new}]{\includegraphics[width = 0.32\textwidth]{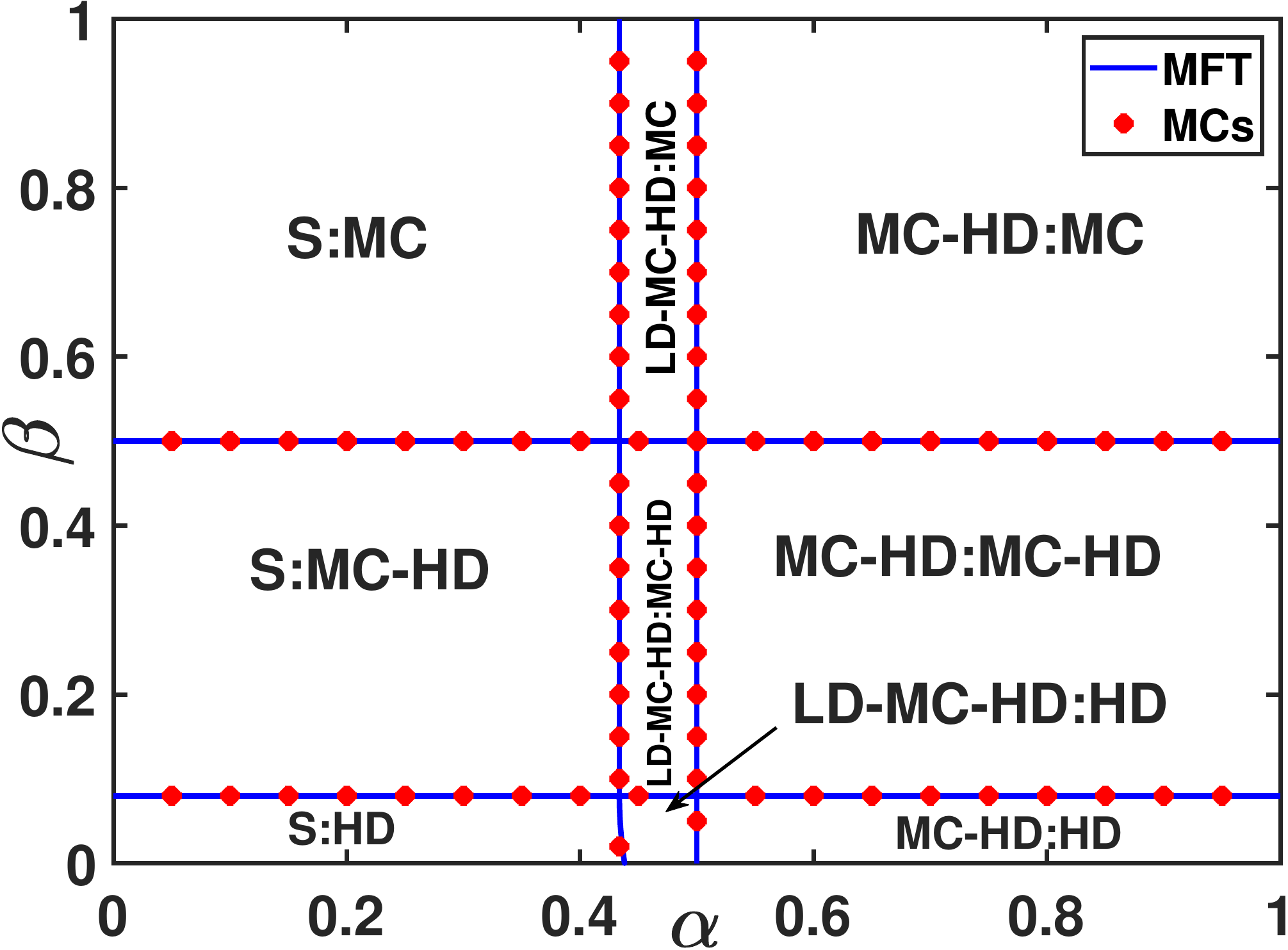}}\\
		\subfigure[\label{fig:Omega=0.5}]{\includegraphics[width = 0.32\textwidth]{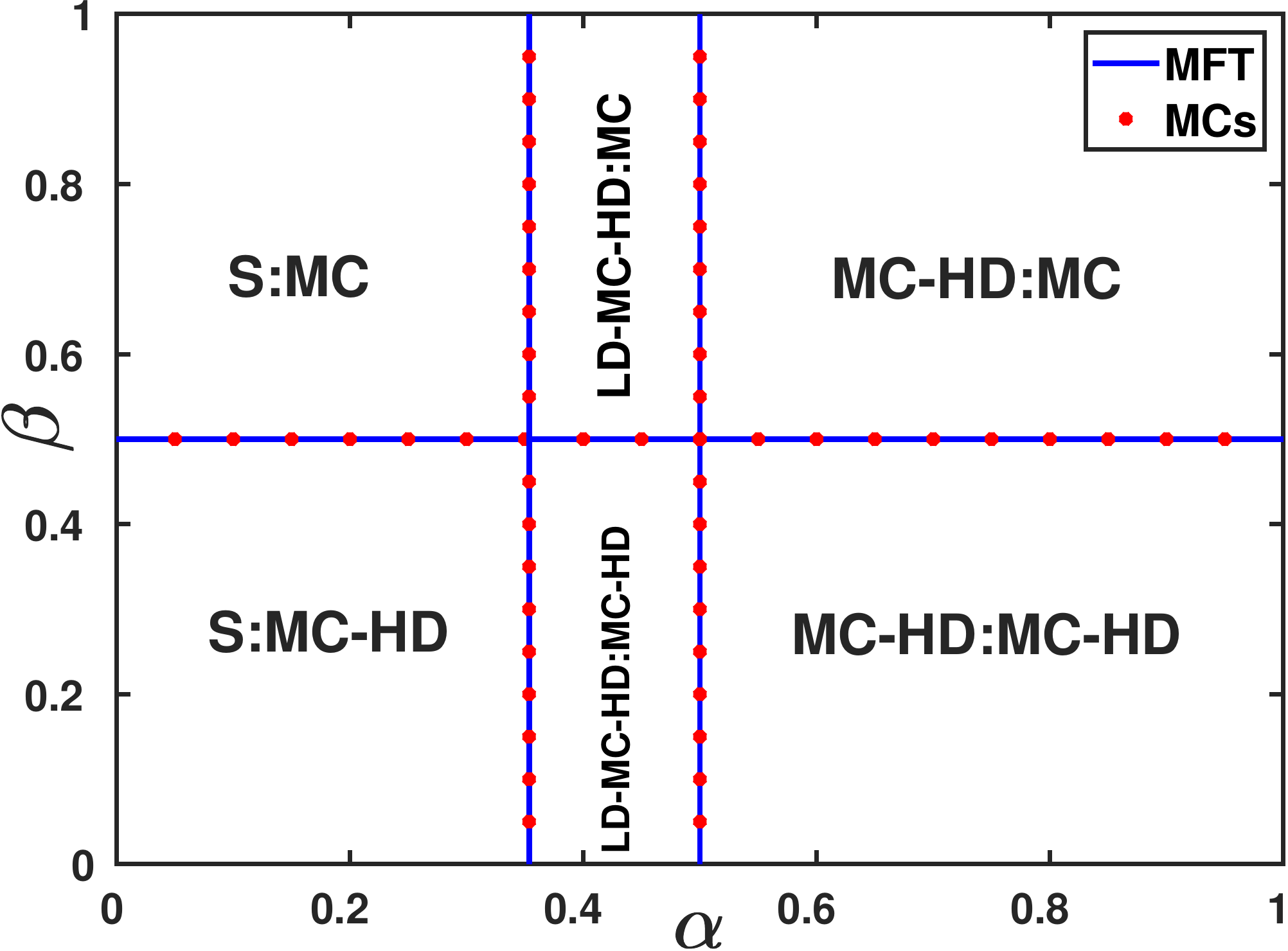}}\quad \quad
			\subfigure[\label{fig:Omega=0.8new}]{\includegraphics[width = 0.32\textwidth]{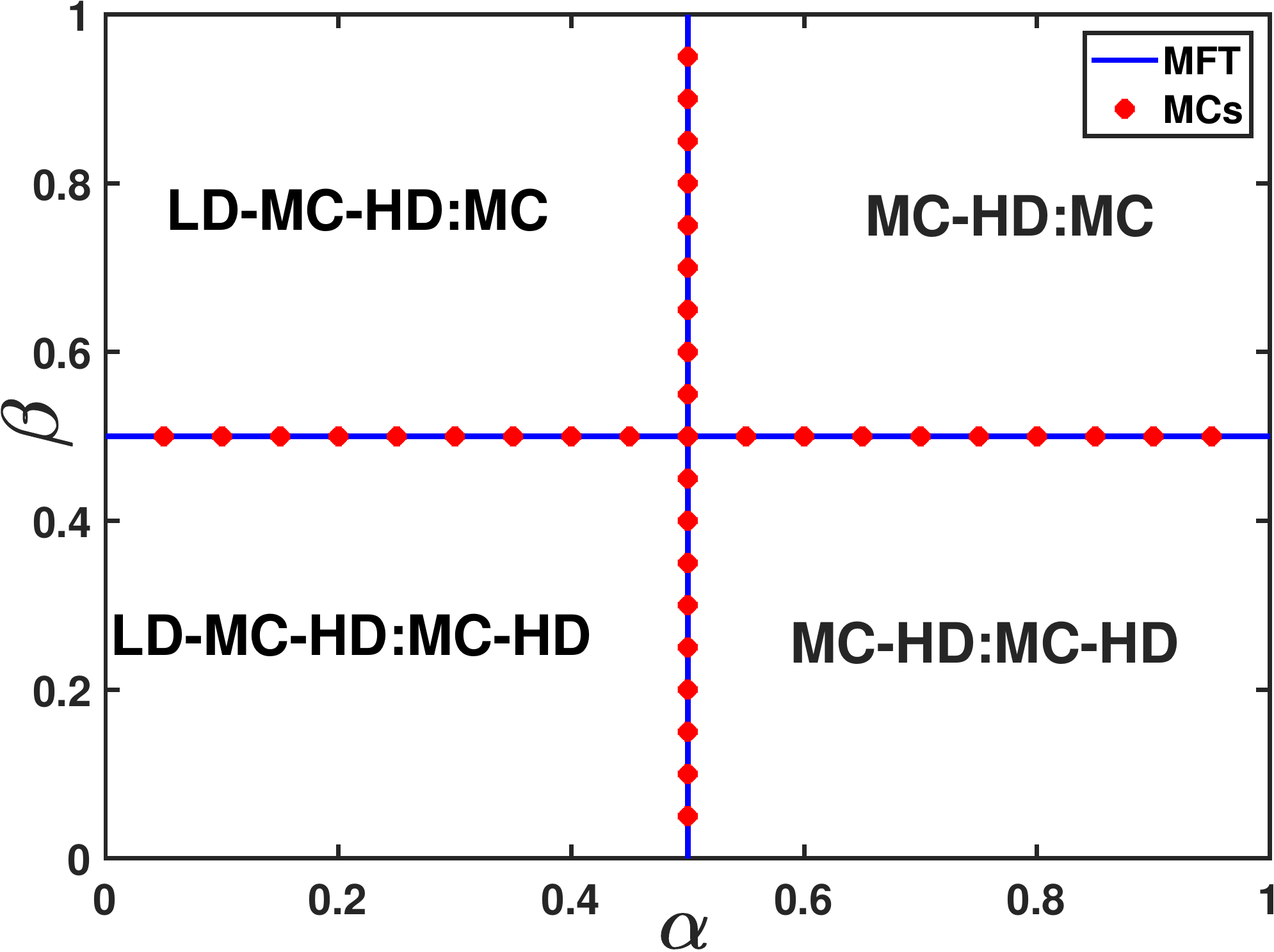}}
	\caption{\label{phase1}Stationary phase diagrams for different value of $\Omega$: (a) $\Omega=0$, (b) $\Omega=0.1$, (c) $\Omega= 1-\frac{\sqrt{3}}{2}$, (d) $\Omega=\frac{1}{2} \left(1-\frac{1}{\sqrt{2}}\right)$, (e) $\Omega=\frac{1}{2\sqrt{2}}$, (f) $\Omega=0.42$, (g) $\Omega=0.5$ and (h) $\Omega=\frac{1}{2} \left(1+\frac{1}{\sqrt{2}}\right)$  for $m=2$ and $n=1$ . The phase transformations are continuous for boundaries between all observed phases. Solid blue lines represents theoretical results and dotted red symbols correspond to Monte Carlo simulation (MCs).}
\end{figure*}

Now let us investigate the case when the number of segments in each subsystem are unequal and for simplicity, we choose $m=2$ and $n=1$ which can be generalized for others values of $m$ and $n\ (m>n)$.   In absence of the non-conserving dynamics $(\Omega=0)$, this case has been well studied \cite{pronina2005theoretical} and for the sake of completeness, we have reproduced its phase diagram as presented in figure \ref{fig:Omega=0}. The phase diagram exhibits $5$ distinct stationary phases, namely, LD:LD, LD:HD, LD:MC, HD:MC and HD:HD. To investigate the effect of LK dynamics on the network, we study the phase diagram by varying $\Omega$ in the parameter space of $\alpha-\beta$. As soon as LK dynamics is introduced in the system, even for a very small value of $\Omega$, the phase composition of the stationary phase diagram is strongly modified. Eight new phases emanate in the system: LD:S, S:HD, LD:LD-MC-HD, S:MC-HD, HD:MC-HD, LD:LD-MC, LD:MC-HD and S:MC along with the five pre-existing phases. For convenience, we have shown the phase diagram for $\Omega=0.1$ capturing these thirteen phases in figure \ref{fig:Omega=0.1new}. 

Further increasing $\Omega$ from $0.1$, results in the shrinkage of LD:LD, LD:S and LD:HD phase and the expansion of S:MC, S:MC-HD and S-HD phases. Besides, no significant change is observed in the topology of the phase diagram till a critical value $\Omega_{C_3}=1-\sqrt{3}/2$, as computed from equation \eqref{LDconditions} at which LD:LD phase disappears. To demonstrate the phase schema at this critical value, we have plotted the phase diagram as shown in figure  \ref{fig:Omega=0.133} for $\Omega_{C_3}=1-\sqrt{3}/2$.

With the increase in  the value of $\Omega$ from $\Omega_{C_3}$, the phase boundary of LD:MC phase approaches the $\beta$ axis and at a critical value $\Omega_{C_4}=(1/2)(1-1/\sqrt{2})$, the feasibility of LD phase in all the incoming segments is disrupted. As a consequence, the possibility of the phases LD:LD-MC, LD:MC, LD:LD-MC-HD, LD:MC-HD, LD:S and LD:HD completely vanishes. Now, the phase diagram becomes simplifier and consists of only six phases namely, S:MC, S:MC-HD, S:HD, HD:MC, HD:MC-HD, and HD:HD as shown in figure \ref{fig:Omega=0.1464new} for $\Omega_{C_4}=(1/2)(1-1/\sqrt{2})$.

Considering the further increase in $\Omega$ from $\Omega_{C_4}$ to $\Omega_{C_5}=1/2\sqrt{2}$, the phases with HD $\rightarrow$ MC phase transition at the junction no longer exists. The existence of such phases requires $\Omega+\beta_{eff}<0.5$ which is only valid when $\Omega<\Omega_{C_5}$. As a result, the LK dynamics dominates and the HD $\rightarrow$ MC phase transition  changes to MC-HD $\rightarrow$ MC phase transition i.e., HD:MC, HD:MC-HD and HD:HD phases transform to MC-HD:MC, MC-HD:MC-HD and MC-HD:HD in the system as presented in figure \ref{fig:Omega=0.356} for $\Omega_{C5}=1/({2\sqrt{2}})$.

Beyond $\Omega_{C_5}$, the $3$-phase coexistence region LD-MC-HD in all the incoming segments is observed in the system. Three new phases LD-MC-HD:MC, LD-MC-HD:MC-HD and LD-MC-HD:HD emanates in the system as illustrated in figure \ref{fig:Omega=0.42new} for $\Omega=0.42$.   As $\Omega$ reaches the value $\Omega_{C_6}=0.5$, the HD phase is no longer observed in all the $R_k$ segments as the phase boundaries of S:HD, LD-MC-HD:HD and MC-HD:HD phases tends to the line $\beta=0$. The phase diagram at the critical value $\Omega_{C_6}=0.5$ is shown in   figure \ref{fig:Omega=0.5}. Now the region for LD-MC-HD:MC and LD-MC-HD:MC-HD phases expand and as a result the phases S:MC and S:MC-HD perishes at the value $\Omega_{C_7}=(1+\sqrt{2}/2)/2$, as plotted in figure \ref{fig:Omega=0.8new}. Finally, after this critical value $\Omega_{C_7}$, increase in $\Omega$ do not produce any topological change in the phase boundaries nor the phase diagram and it remains unaltered even for $\Omega \to \infty$. Hence, the phase diagram remains invariant and portrays only four phases namely, LD-MC-HD:MC, LD-MC-HD:MC-HD, MC-HD:MC and MC-HD:MC-HD. Clearly, the stationary phase diagrams show non-monotonic behaviour with respect to the equal attachment/detachment rate $\Omega$ on the number of phases in the $V(2,1)$ network.

Now, we analyse the possible phase transitions that occur for a fixed $\Omega$ and study the phase diagram for $\Omega=0.1$. Choosing fixed $\beta \in (\alpha_{eff} -\Omega, \alpha_{eff}+\Omega)$, where $\alpha_{eff}$ is given by equation \eqref{LDeff} depending upon $\alpha$ and $\Omega$, we can clearly visualize that  when $\alpha$ is increased, the phase changes from LD:S to LD:HD then to S:HD and finally to HD:HD (figure \ref{fig:Omega=0.1new}). This behaviour can be easily understood by the following arguments. With an increment in $\alpha$, $\alpha_{eff}$ increases and, hence the influx of particles increases in the outgoing segments. Consequently, the shock vanishes from all the outgoing segments and HD phase emerges in all these segments. Further, increasing $\alpha$, the values of $\alpha_{eff}$ and $\beta_{eff}$ (equation \eqref{HDeff}) no longer varies which means that the system dynamics will now be governed by only varying parameter $\alpha$ in $L$ subsystem and thus, S:HD phase emerges in the system. This is because, with the increase in the entry rate, the junction-induced shock absorbs the incoming particles and travels towards the left side in all the incoming segments. Finally, with further increase in the entry rate, the HD phase is observed in all the segments of the network. Similarly, for any fixed $\beta \in (0,1)$ and varying $\alpha$, an analogous argument holds for the phase transitions in the system. 

Though in the above discussion, our analysis was restricted to the case when the number of segments in the $L$ subsystem is greater than the number of segments in the $R$ subsystem $(m>n)$. For $m<n$, the phase transitions occurring at the junction and the corresponding phase diagram can be examined utilising the particle-hole symmetry. The system exhibits particle-hole symmetry as advancing particles corresponds to receding holes. A particle entering through the first site of ${L_k}'s$ can be interpreted as a hole leaving these segments and vice versa for the last site of ${R_k}'s$ segments. Analogously, attachment/detachment of particles can be mapped to detachment/attachment of holes  in the bulk. Therefore, utilising the following transformations, the case when $m<n$ can be easily explained on similar lines as for the case $m>n$,
\begin{align*}
 \rho &\leftrightarrow 1- \rho\\
 \Omega_A&\leftrightarrow \Omega_D\\
 \alpha&\leftrightarrow \beta\\
n &\leftrightarrow  m\\
 x&\leftrightarrow -x\\
 \alpha_{eff}&\leftrightarrow \beta_{eff}.\\
  \end{align*}
In the above discussion, our analysis was restricted to the case when the number of segments in the $L$ subsystem is greater than the number of segments in the $R$ subsystem $(m>n)$.  For $m<n$, the phase transitions occurring at the junction and the corresponding phase diagram can be examined on similar lines. Moreover, the phase regimes that exist for different values of $m$ and $n$ are described in tabular form in Table \ref{change}.  
\begin{table}[htb]
\begin{center}
\caption{The possible phase transitions that can occur across the junction for different values of $m$ and $n$.}

\begin{tabular}{||c|| c|c|c||} 
 \hline \hline
Phase Transition at the junction & $m>n$ & $m=n$ & $m<n$  \\ 
 \hline \hline
 LD $\rightarrow$ LD & $\surd$   &$\surd$  &$\surd$ \\ \hline
 LD $\rightarrow$ MC & $\surd$   &$\times$  &$\times$ \\ \hline
 LD $\rightarrow$ HD & $\surd$   &$\surd$  &$\surd$ \\ \hline
MC$\rightarrow$ LD & $\times$   &$\times$  &$\surd$ \\ \hline
MC $\rightarrow$ MC & $\times$   &$\surd$  &$\times$ \\ \hline
MC $\rightarrow$ HD & $\times$   &$\times$  &$\surd$ \\ \hline
HD $\rightarrow$ LD & $\times$   &$\times$  &$\times$ \\ \hline
HD $\rightarrow$ MC & $\surd$   &$\times$  &$\times$ \\ \hline
HD $\rightarrow$ HD & $\surd$   &$\surd$  &$\surd$ \\ \hline
 \hline
\end{tabular}
\label{change}
\end{center}
\end{table}

\section{\label{conclude}Conclusion}

To summarize, we have presented a detailed study of a $V(m:n)$ network consisting of $m$ incoming segments and $n$ outgoing segments connected via a junction. This network is equipped with an additional feature of particle-creation and annihilation where a particle can bind to an empty site or unbind from an occupied one with given rates. Our theoretical method is based on the idea that each segment can be viewed as a one dimensional TASEP incorporated with LK. This allows us to implement the simple mean field approximation to investigate the crucial steady state properties of the system such as density profiles, phase diagrams and phase transitions. In the support of mathematical investigations, the theoretical outcomes are obtained for all the observed phases for equal attachment-detachment rates. Our findings are theoretically examined for the different number of the incoming and the outgoing segments and numerically by extensive Monte Carlo simulations.

We specifically consider two distinct scenarios for the system dynamics: when the number of segments in both the subsystems is different and when these values are the same. The study reports the explicit expressions for the phase boundaries of all the possible feasible phases and also provided valid arguments for the non-existence of certain phases in the system. Among the $49$ possible phases,  $19$ different stationary phases have been observed in the system for varied values of $\Omega$ when $m\neq n$. For the case when $m=n$, the number of perceived phases reduces to $13$. The analysis found that the system displays a large number of  stationary phases, so for a systematic study, the potential phases have been divided into various subclasses based upon the dynamics happening at the junction. We observed that when the number of incoming and outgoing segments are equal, the maximal current phase can persist in all the segments.

Further, we study the effects of LK rates and the number of segments in each subsystem on the system dynamics. It has been seen that the effect of $\Omega$ on the number of phases is non-monotonic. Introducing the LK dynamics in the system, for $m\neq n$, firstly increases the number of phases from $4$ to $13$, which reduces to $6$, then further rises to $9$ and finally decreases to $4$ as $\Omega$ increases. Whereas for $m=n$, with an increase in $\Omega$, the number of phases escalates from $3$ to $13$ and then drops to $4$ phases. After a certain value of $\Omega$, the topology of the phase diagram remains unaltered and the phases boundaries, as well as the observed phases, are no longer modified and hence, the number of observed phases remains $4$ as $\Omega \to \infty$. The critical values of $\Omega$ are computed where the appearance or disappearance of phases is observed. We also found that when the number of incoming and outgoing segments are equal, the obtained phase diagram is the analog of a single segment TASEP-LK model \cite{parmeggiani2004totally, parmeggiani2003phase}.

The proposed model is an attempt to provide an intrinsic means to interpret the steady state properties of transport phenomenon on roads, molecular filaments, etc under the influence of LK dynamics. A further extension of this study can deal with the scenario where particle exchange is permitted in both subsystems. Our results can also be extended to networks with various junctions.
  \\

  \noindent
\textbf{Acknowledgments}\\

The first author thanks Council of Scientific \& Industrial Research (CSIR), India for financial support under
File No: 09/1005(0023)/2018-EMR-I. AKG acknowledges support from DST-SERB, Govt. of India (Grant
CRG/2019/004669).\\

\end{document}